# Can AI Solve the Peer Review Crisis? A Large-Scale, Cross-Model Experiment of LLMs' Performance and Biases in Evaluating over 1,000 Economics Papers

Pat Pataranutaporn

*Massachusetts Institute of Technology, Cambridge, MA, USA*

Nattavudh Powdthavee

*Nanyang Technological University, Singapore*

Chayapatr Achiwaranguprok

*University of Thai Chamber of Commerce, Thailand*

Pattie Maes

*Massachusetts Institute of Technology, Cambridge, MA, USA*

## Abstract

This study examines the potential of large language models (LLMs) to augment the academic peer review process by reliably evaluating the quality of economics research without introducing systematic bias. We conduct one of the first large-scale experimental assessments of four state-of-the-art LLMs—GPT-4o, Claude 3.5, Gemma 3, and LLaMA 3.3—across two complementary experiments. In the first, we use nonparametric binscatter and linear regression techniques to analyze over 29,000 evaluations of 1,220 anonymized papers drawn from 110 economics journals excluded from the training data of current LLMs, along with a set of AI-generated submissions. The results show that LLMs consistently distinguish between higher- and lower-quality research based solely on textual content, producing quality gradients that closely align with established journal prestige measures. Claude and Gemma perform exceptionally well in capturing these gradients, while GPT excels in detecting AI-generated content. The second experiment comprises 8,910 evaluations designed to assess whether LLMs replicate human-like biases in single-blind reviews. By systematically varying author gender, institutional affiliation, and academic prominence across 330 papers, we find that GPT, Gemma, and LLaMA assign significantly higher ratings to submissions from top male authors and elite institutions relative to the same papers presented anonymously. These results emphasize the importance of excluding author-identifying information when deploying LLMs in editorial screening. Overall, our findings provide compelling evidence and practical guidance for integrating LLMs into peer review to enhance efficiency, improve accuracy, and promote equity in the publication process of economics research. (242 words).

**Keywords**: Artificial intelligence; peer review; large language model (LLM); bias in academia; economics publishing; binscatter regression

**JEL:** A11, C63, O33, I23

## 1. Introduction

The economics discipline has long grappled with two persistent challenges in the peer review process: an insufficient supply of willing and qualified referees and prolonged turnaround times, which can extend to two years or more (Ellison, 2002; Card & DellaVigna, 2013). These issues are magnified by the outsized career implications of publishing in the discipline's "top five" journals—*American Economic Review*, *Quarterly Journal of Economics*, *Journal of Political Economy*, *Econometrica*, and *Review of Economic Studies* (Heckman & Moktan, 2020). To address these challenges, several reforms have been introduced, including shorter article formats (e.g., "Insights" or "Short Papers"), limitations on revise-and-resubmit rounds, and monetary compensation for referees.

While these efforts have led to some improvements, progress has been more incremental than transformative. Leading journals like *Econometrica* and the *Review of Economic Studies*, which attract disproportionately high submission volumes (Card & DellaVigna, 2013), still face significant challenges in recruiting referees. These journals have median first decision times of 3 to 6 months, which is significantly longer than in psychology (2 to 4 months) and political science (3 to 4 months). This inefficiency creates burdens on junior economists, who must navigate strict tenure requirements demanding multiple publications in top-tier journals within tight timeframes. Consequently, delays in the review process compound the already considerable pressures faced by early-career researchers.

Recent advances in large language models (LLMs)—both closed-source, such as OpenAI's GPT and Anthropic's Claude, and open-source, like Google's Gemma and Meta's LLaMA—have sparked debates about whether artificial intelligence (AI) could help address the so-called "peer review crisis" by easing the workload on human referees and optimizing editorial workflows (Yuan et al., 2022; Mehta et al., 2024; Saad et al., 2024). Nevertheless, the potential role of AI in peer review raises significant concerns regarding reliability, transparency, and bias (Bender & Koller, 2020). Among these concerns is whether AI can consistently assess the quality of academic work, effectively distinguishing between high-quality, medium-quality, and low-quality submissions and those generated by AI itself.

A secondary but crucial question is whether AI systems replicate the behavioral biases typically observed in human reviewers. Evidence from economics shows that referees and editors, consciously or unconsciously, often favor authors with well-known reputations or connections to prestigious institutions (Blank, 1991; Brogaard et al., 2014; Huber et al., 2022). Given that AI models are trained on a vast array of human-generated text, which frequently includes author identities and affiliations from publicly available papers, these systems may perpetuate or even amplify existing biases. Whether AI-based reviewers reduce biases by focusing solely on textual content or exacerbate them by incorporating patterns learned from broader datasets remains an open and largely unexplored question.

This study is the first to systematically evaluate the performance of both commercial and open-source large language models (LLMs)—including GPT-4o mini, Claude 3.5 Haiku, Gemma 3 27B, and LLaMA 3.3 70B—in assessing the quality of 1,220 papers. We began by compiling a recent sample of articles from 100 economics journals. For each journal, we randomly selected 10 articles that had not yet been incorporated into the AI training corpus and removed identifying information about the authors and the journals before presenting them to the LLMs. We then compared the LLMs' evaluations of each paper with the journal's ordinal ranking in the IDEAS/RePEc aggregate rankings of economics journals, as reported on the RePEc/IDEAS website. In effect, this process allows us to compare each LLM's subjective assessment of paper quality with an objective benchmark—the quality of the journal in which the paper was published, based on aggregated metrics such as impact factor and h-index—thereby enabling us to assess whether LLMs can distinguish a paper's objective quality as effectively as journal editors.

As part of our assessment of LLMs' capabilities, we compare 1,000 papers drawn from RePEc-ranked journals with two additional sets: (1) 100 papers from 10 paid open-access journals not ranked within RePEc, and (2) 120 AI-generated fake papers. The fake papers are created using both closed-source models—OpenAI o1 (reasoning model), OpenAI GPT-4o, and Anthropic Claude 3.7 Sonnet with Extended Reasoning—and open-source models—DeepSeek-R1 (reasoning model), Gemma 3 27B, and LLaMA 3.1 405B. We then prompt the LLMs to evaluate each paper's quality and assess the likelihood that it was AI-generated. This analysis allows us to test whether LLMs can distinguish not only between papers from ranked and unranked journals, but also between genuine research and AI-generated papers produced by a range of different models.

Additionally, given that most economics journals follow a single-blind review policy—where author identities are known to editors and referees—we examine whether LLMs exhibit biases against minority or lesser-known authors. To do so, we systematically vary author characteristics—such as gender (e.g., top-ranked male and female economists from RePEc's top 10 list), professional status (e.g., lower-ranked economists), and name type (e.g., randomly generated names)—as well as institutional affiliations, across a subset of 30 papers taken from 3 journals of varying quality. Findings from this experimental design (N = 8,100 unique submissions) will help determine whether author information in submitted manuscripts should be disclosed to or withheld from AI systems during evaluation.

Our findings indicate that all LLMs reliably distinguish papers by overall quality, as well as by dimensions such as originality, rigor, scope, and impact, across low-, medium-, and high-ranked economics journals. Among them, OpenAI's GPT and Google's Gemma demonstrate the strongest evaluative gradients across journal tiers, while Meta's LLaMA exhibits the least sensitivity to changes in journal ranking. We provide robust evidence that LLM-assigned scores decline approximately linearly with journal rank, decreasing steadily from top-tier to lower-tier publications. These results remain robust after controlling for paper length and conducting a range of robustness checks. Notably, these patterns persist even within subfields. For instance, within finance, Gemma assigns higher average ratings to papers published in the *Journal of Finance* (ranked 7$^{th}$) than to those in the *Journal of*

*Corporate Finance* (ranked 42nd). A similar pattern emerges within applied micro-econometrics, where the *American Economic Journal: Applied Micro* (ranked 10th) receives higher scores than the *Journal of Health Economics* (ranked 38th).

We also find consistent evidence that all LLMs can reliably distinguish between papers published in RePEc-ranked journals and those in unranked outlets. However, their ability to detect fake, AI-generated papers varies considerably. The most effective detection strategy combines Gemma's quality ratings with GPT's assessments of whether a paper was written by AI. When we differentiate between fake papers generated by different LLMs—both reasoning models (OpenAI o1, DeepSeek-R1, Anthropic Claude 3.7 Sonnet with Extended Thinking) and non-reasoning models (OpenAI GPT-4o, Gemma 3 27B, LLaMA 3.1 405B)—we find that those produced by Claude are, on average, the most likely to receive the highest quality ratings and the least likely to be identified as fake by other LLMs.

Despite their ability to assess quality, our additional experiment finds that three out of four LLMs exhibit bias favoring well-known authors, whereas all four LLMs shows bias toward authors from prominent institutions such as MIT—even when the quality of the evaluated paper is held constant. This evidence of AI bias indicates that if editors plan to integrate LLMs into the manuscript screening process, they should ensure that author names, affiliations, and potentially other identifying characteristics—such as gender and ethnicity—are withheld to reduce the risk of biased evaluations.

The remainder of this paper is organized as follows: Section 2 reviews the literature on biases in peer review and the emerging role of AI in academic publishing, highlighting key gaps in understanding AI's impact on equity and efficiency. Section 3 introduces our methodology and empirical strategy, detailing the design and data analysis. Section 4 presents the results, and Section 5 offers theoretical implications based on the paper's empirical findings. Section 6 discusses and concludes with recommendations for future research and policy.

## 2. Background Literature

### 2.1. Inefficiencies and Biases in Editorial and Peer Review Decisions

The inefficiency of the publishing process in economics is well-documented. Early empirical work by Ellison (2002) reveals that in the 1970s, only about 20–25% of accepted papers in top-tier economics journals underwent two or more rounds of peer review. By the 1990s, this figure had approximately doubled to 40–50%, indicating a significant increase in the interactions between authors, referees, and editors. He also noted a rise in the median time to acceptance at the *Journal of Political Economy*, which grew from 4–6 months in the early 1970s to 9–12 months by the 1990s. This notable increase in review times may be attributed to more stringent demands for robustness checks by referees and increased competition for space in top journals, both of which have led to multiple rounds of revisions as a method of filtering submissions. Another contributing factor to the lengthy review process is the moderate rate at which potential referees decline editors' invitations to review manuscripts. For instance, Hamermesh (1994) reports that approximately 17% of invited referees declined the request, while an additional 5% of those who accepted either took more than six months to complete the review or failed to submit a report altogether.

Given the heightened competition for publication in top economics journals since Ellison's (2002) study, along with the persistently low incentives for potential reviewers to accept editorial requests, as highlighted by Hamermesh (1994), the turnaround times reported in earlier decades have likely increased significantly over time. The lengthening of the review process also applies to the duration required for the initial editorial response. For instance, Azer (2007) provides evidence that the average first response time in economics journals has risen from about 2 months to between 3 and 6 months over the past four decades.

Using submission and publication data from the "top five" economics journals, Card and DellaVigna (2013) also documented a substantial rise in annual submissions—from around 2,800 in the early 1990s to over 5,800 by 2012. However, despite the sharp increase in submissions, the total number of articles published by these journals declined—from approximately 400 per year in the late 1970s to 300 or fewer by the early 2010s. This suggests that many high-quality papers that meet the standards of a top-five journal are likely rejected, mainly due to the sheer volume of submissions that editors must manage each year.

Beyond evidence of inefficiencies in the economics publishing process, a substantial body of research has examined how factors such as gender, institutional affiliation, and the prominence of "star" authors influence editors' assessments of the quality of submissions to top journals. One of the earliest systematic studies, conducted by Blank (1991), examined the effects of transitioning from single- to double-blind reviewing at the *American Economic Review* (AER). The study found that while the shift to double-blind review did not drastically change overall acceptance rates, it modestly improved outcomes for female economists and authors from lower-ranked institutions. Blank's work provides some of the earliest evidence that revealing an author's identity can introduce biases favoring well-known scholars or those affiliated with prestigious institutions. However, her findings are limited by the specific context of a single journal and may now be less applicable due to changes in the discipline over the years. For instance, most economics papers are now available as pre-prints, making it likely that referees are already aware of the authors' identities and affiliations, thereby undermining the effectiveness of double-blind review.

Given that most, if not all, papers submitted to economics journals over the last few decades are reviewed under a single-blind process, a significant strand of research has focused on whether female economists face systematic disadvantages in the peer-review process. Using data from four leading journals, Card et al. (2020) find little evidence of outright discrimination at the decision stage once relevant article- and author-level variables are controlled. Female-authored submissions are not overtly penalized in terms of immediate acceptance or rejection. However, differences emerge when examining revise-and-resubmit (R&R) invitations: the authors observe modest, albeit not always statistically robust, disparities of around 1.7%, suggesting that women may be slightly less likely than men to receive an R&R in borderline cases.

Other studies highlight more subtle forms of gender bias. Hengel (2022), for instance, documents that female-authored papers often endure longer review times (around 3-6 months) and receive more exhaustive feedback on writing style and clarity. This pattern, sometimes characterized as a "higher bar," may reflect referees' unconscious assumptions about female competence or writing quality. Even if the ultimate acceptance rate is not lower, the cumulative effect of extensive revisions and protracted timelines can hamper female scholars' publication records and career progression. Such subtle biases may not be easily captured by simple acceptance rate comparisons, indicating that journals and editorial boards should examine both *how* authors are reviewed and *how long* each step in the process takes.

Another strand of literature examines the influence of institutional rankings and prominent authors on editorial and peer review decisions. In a seminal paper, Laband and Piette (1994) investigate whether higher acceptance rates for authors affiliated with journal editors reflect favoritism or a genuine effort to select superior work. By analyzing publication outcomes and subsequent citation metrics, they find that editor-affiliated authors are more likely to have their papers accepted. However, these papers also tend to perform better in terms of citations. This dual finding complicates the interpretation of editorial bias, suggesting that while personal connections may confer an advantage, the research produced by editor-affiliated authors often demonstrates substantial impact.

Another notable study is by Card and DellaVigna (2020). While their primary focus is on how *AER* editors incorporate citation prospects and referee recommendations into their decision-making, they also identify a significant influence of institutional affiliations on revise-and-resubmit (R&R) outcomes. However, like Laband and Piette (1994), the authors caution that it is challenging to disentangle the effects of institutional affiliation from the author's reputation, as researchers at well-resourced institutions often have greater capacity to produce more polished and innovative work.

More recently, Huber et al. (2022) investigated the impact of author prominence on peer review outcomes through a preregistered field experiment. They submitted a finance manuscript co-authored by a Nobel laureate and a relatively unknown early-career researcher to 3,300 potential reviewers, varying the visibility of the author names. Their findings revealed a strong status bias: when the prominent author's name was shown, the manuscript received significantly fewer rejection recommendations (22.6%) compared to the anonymized version (48.2%) and the version attributed to the less prominent author (65.4%). The bias extended to more favorable overall assessments of the manuscript's quality. However, it remains to be seen whether the large effect size observed in this study will be replicated in real editorial decisions, where the stakes are significantly higher.

Given the documented biases associated with editorial and peer review decisions, coupled with challenges such as the scarcity of willing, high-quality referees and prolonged turnaround times for economics journals (Ellison, 2002; Azer, 2007), there is a clear need for a new and systematic approach to peer review within the discipline.

**2.2. Artificial intelligence (AI) as a screener and reviewer**

One promising avenue is the integration of artificial intelligence (AI) as a supplementary tool in the review process. One hypothesis is that AI systems could assist by evaluating technical rigor, checking for methodological consistency, identifying potential errors, and even providing initial assessments of the manuscript's contribution based on citation patterns and relevance to existing literature.

There is currently a small but growing body of research within computer science investigating AI's potential to enhance and improve the peer-review process. One notable example is Yuan et al. (2022), who explore the possibility of natural language processing (NLP) systems to generate comprehensive and aspect-sensitive peer reviews for scientific papers. Using a dataset of machine learning papers annotated with aspect-based review information, they train and evaluate NLP models capable of producing comprehensive, aspect-sensitive review drafts. The study reveals that while the models can accurately summarize a paper's core ideas and provide broader aspect coverage than human reviewers, their reviews are often non-factual and lack constructive criticism. They conclude that although the current technology is not yet ready to replace human reviewers, it has potential as a tool to assist reviewers and authors in identifying key strengths and weaknesses in manuscripts. While the paper finds some evidence of bias against non-native speakers in terms of clarity and perceived potential impact, it does not address biases related to affiliation, gender, or author prominence.

In another study, Checco et al. (2021) investigate the potential of AI to enhance the peer-review process by predicting review outcomes and identifying biases. Using a neural network model trained on peer review data from three academic conferences, the study incorporates features such as word distributions, readability metrics, and document formatting to predict reviewer decisions. The results indicate that AI can predict review outcomes with significant accuracy, suggesting that superficial features like formatting and readability correlate with reviewer judgments. However, the study raises concerns about algorithmic bias, as AI systems may reinforce biases already present in human reviewers, such as those related to language, regional representation, and first impressions. Like Yuan et al. (2022), they also caution that AI is not yet suitable to replace human reviewers. Nevertheless, they propose that editors and reviewers can use AI to pre-screen tasks, identify systematic biases, and improve the efficiency of the review process.

Despite recent advancements in AI and its potential applications in peer review, existing studies have predominantly relied on observational data rather than employing randomization to establish causal relationships. Consequently, a significant gap remains in understanding whether AI systems treat identical papers differently based on authors' affiliations, gender, or prominence. Furthermore, previous studies have not utilized already published papers that were previously evaluated by human reviewers, making it difficult to directly compare AI-generated ratings with human judgments of a paper's overall quality. For example, it is unclear whether AI can reliably distinguish between papers accepted in top-tier journals versus those published in mid-tier journals. Addressing these limitations requires systematic research to benchmark AI systems against human assessments using papers with known publication outcomes, offering a clearer understanding of LLM's capabilities and potential biases in peer review. To address these limitations, this paper is one of the first in both economics and computer science to take an experimental approach to provide a rigorous evaluation of AI's capabilities and biases in the peer-review process within the field of economics.

## 3. Methods

### 3.1. Data

We pre-registered our main study on AsPredicted.org (https://aspredicted.org/24qd-8t4w.pdf). To assess the capacity of large language models (LLMs) to evaluate economics research, we manually downloaded 1,000 recently published or Online First articles—10 per journal—from 100 economics journals ranked on the RePEc website (https://ideas.repec.org/top/top.journals.all.html), covering publications from late 2024 to early 2025. In selecting these 100 journals, we excluded review journals such as the *Journal of Economic Literature* and the *Journal of Economic Perspectives*, short-format outlets like *Economics Letters*, proceedings volumes such as those published by the Federal Reserve Bank of San Francisco, and journals clearly outside the economics discipline, such as the *Strategic Management Journal*. In addition, we compiled a sample of 100 published articles from for-profit, open-access academic publishers that are not ranked on RePEc. We intentionally selected only recently published papers to ensure that the data had not yet been incorporated into the latest versions of AI systems, thus preventing LLMs from having prior knowledge of where the papers were published or who the authors were.[1]

We employed three open-source models—Deepseek-R1 (reasoning model), Gemma 3 27B, and LLaMA—and three closed-source models—o1 (reasoning model), GPT-4o mini, and Claude 3.5 Haiku with extended reasoning capabilities—to generate five papers in each of four broad fields: Macroeconomics, Microeconomics, Applied Micro-econometrics, and Finance. In total, this produced 120 fake, AI-generated papers. The average paper contains 12,334.53 words, with a standard deviation of 6,273.99 and a median of 12037.5. Word counts range from a minimum of 326 to a maximum of 58,925. In comparison to human-written articles, AI-generated papers are significantly shorter, averaging just 1,427.32 words. In terms of economic subfields, 320 papers were published in applied micro-econometrics journals, 110 in finance journals, 220 in general economics journals, 170 in macroeconomics journals, 120 in microeconomics journals, and 280 in journals covering other subfields. Appendix A provides a list of the journals included in the analysis.

[1] The following are the cut-off dates for the training data for each of the four LLMs: Llama 3.3 70B (December 1, 2023: https://huggingface.co/meta-llama/Llama-3.3-70B-Instruct), GPT-4o mini (November 2023 to June 2024: https://help.openai.com/en/articles/9624314-model-release-notes), Claude 3.5 Haiku (July 2024: https://docs.anthropic.com/en/docs/about-claude/models/all-models), Gemma 3 27B (March 2024: https://gradientflow.com/gemma-3-what-you-need-to-know/). All the downloaded RePEc-ranked papers, along with most unranked papers—particularly those published in MDPI journals—were published between late 2024 and early 2025.

We then input the full text of all 1,220 papers—excluding any information related to the publishing journal, as well as author names and affiliations for the human-written papers[2]—into four large language models for evaluation: two closed-source models (GPT-4o mini and Claude 3.5 Haiku) and two open-source models (Gemma 3 27B and LLaMA 3.3 70B).[3] For the primary outcome variable, each paper was evaluated three times using the following prompt: "In your capacity as a reviewer for one of the most prestigious and highly selective top-five economics journals (such as Econometrica, Journal of Political Economy, or the Quarterly Journal of Economics), please determine whether you would recommend this submission for publication using the following 6-point scale:

- 1 = *Definite Reject*: Fatal flaws in theory/methodology, insufficient contribution, or serious validity concerns that make the paper unsuitable for the journal.
- 2 = *Reject with Option to Resubmit*: Significant issues with theory, methodology, or contribution, but potentially salvageable with major revisions and fresh review.
- 3 = *Major Revision*: Substantial changes are needed to theory, empirics, or exposition, but the core contribution is promising enough to warrant another round.
- 4 = *Minor Revision*: Generally strong paper with a few small changes needed in exposition, robustness checks, or literature discussion.
- 5 = *Very Minor Revision*: Excellent contribution, needing only technical corrections or minor clarifications.

---

[2] Details of the methods we used to exclude all journal and author information, as well as the data and codes used in this study, can be found here:: https://github.com/mitmedialab/ai-peer-review-crisis.

[3] Among the LLMs used in this study, Gemma 3 27B, LLaMA 3.1 405B, and GPT-4o mini each have a context window of 128,000 tokens, while Claude Sonnet 3.7 supports up to 200,000 tokens. Given that one token roughly corresponds to 0.75 words, this means these models can process approximately 96,000 words (for the 128K-token models) to 150,000 words (for the 200K-token model) at once—enough to analyze the full text of most academic papers and even multiple papers or full-length books in a single input. We also maintain the default temperature setting, as recommended by the LLM model provider, as optimal for regular tasks.

- 6 = *Accept As Is*: Exceptional contribution ready for immediate publication."

The average of the three evaluations across four LLMs is then calculated and used as the outcome variable. In addition, we prompted the LLMs to evaluate each paper three times on a 10-point scale across five dimensions: originality, rigor, scope, impact, and written by AI. See Appendix B for all the prompts used in this analysis. We conducted 29,280 evaluations in total.

For our analysis of AI bias, we selected a subset of 330 papers across 110 journals—three papers per journal—drawn from both RePEc-ranked and unranked journals. We systematically varied each paper by modifying the authors' characteristics along three key dimensions: gender, institutional affiliation, and academic prominence. For institutional affiliation, each paper was attributed to authors affiliated with one of the following institutions: the Massachusetts Institute of Technology (MIT), Harvard University, the London School of Economics (LSE), the University of Cape Town (South Africa), Nanyang Technological University (Singapore), and Chulalongkorn University (Thailand). MIT, Harvard, and LSE host globally renowned economics departments, whereas the University of Cape Town, NTU, and Chulalongkorn University, while nationally respected, are less internationally recognized for their economics programs.

To introduce variation in gender and academic reputation, we replaced the original authors of the base articles with a new set drawn from two categories: (1) prominent economists—comprising the top five male and top five female economists from the RePEc top 25% list, and (2) non-academic individuals—five male and five female randomly generated names with no academic or professional affiliation. The baseline condition for all categorical variables was a fully anonymized ("blind") version of each paper, in which no information about the authors' names or institutional affiliations was provided. This resulted in a total of 8,910 observations: 6,600 from 330 papers rated under 20 different author names, 1,980 from 330 papers rated under six different institutional affiliations, and 330 from the anonymized (blind) versions of the papers. Unlike the journal quality analysis—where each paper was evaluated three times and the results aggregated to produce a single outcome per paper—each version of a paper with varying author characteristics was evaluated only once by the LLMs.

### 3.2. Empirical Strategy

Our primary empirical strategy examines how LLMs evaluate the quality of academic papers based on the journal in which each paper is published. The dependent variable is the average quality score assigned to each paper across multiple LLMs. The main independent variable is the journal's ranking, determined using RePEc's composite index, where lower values correspond to higher-ranked journals.

The analysis focuses on a sample of 1,000 papers published in journals listed in the RePEc ranking, ensuring a consistent and objectively defined measure of journal quality. This framework allows us to treat journal ranking as a continuous variable and to flexibly estimate how LLM-assessed paper quality varies across the journal quality distribution.

Let $Y_i$ denote the LLM-assigned average quality score for paper $i$, and let $R_j$ denote the RePEc ranking of the journal $j$ in which paper $i$ is published. We are interested in estimating the conditional expectation function:

$$\mu(R_j) = E[Y_i|R_j]. \qquad (1)$$

To estimate $\mu(\cdot)$ flexibly and nonparametrically, we adopt the framework developed by Cattaneo et al. (2024, 2025). We partition the running variable (journal ranking) into 100 fixed bins, each containing exactly 10 papers, sorted by journal ranking. Within each bin, we estimate a linear polynomial approximation to the conditional expectation function. This choice ensures stable estimation given the small number of observations per bin.

To correct for estimation bias and construct valid confidence intervals, we apply a global cubic polynomial fit across the entire range of the running variable. This global polynomial is used to implement bias correction and construct confidence intervals for the estimated conditional expectation function, evaluated at representative values of the ranking variable, typically at the center of each bin. These intervals reflect global inference, rather than uncertainty within individual bins.

The full estimating equation is:

$$Y_i = \mu(R_j) + \beta_1 Length_i + \beta_2 Length_i^2 + \gamma_f + \varepsilon_i, \qquad (2)$$

where $Length_i$ is the word count of paper $I$, $Length_i^2$ allows for non-linear effects of the paper's length, and $\gamma_f$ denotes the journal's field fixed effects (applied micro-econometrics, microeconomics, macroeconomics, general, and finance). The function $\mu(R_j)$ is estimated using piecewise polynomials fit within bins, with bias correction and inference based on the global cubic polynomial.

Since the LLMs evaluated papers without access to journal identity or ranking information, we do not expect systematic correlation in the residuals across published articles from the same journal. Accordingly, we compute heteroskedasticity-robust standard errors, which allow for arbitrary variance in the error term across observations without imposing any cluster structure.

This nonparametric design forms the foundation of our empirical strategy. It leverages the ordinal structure of the RePEc journal rankings to assess whether LLM-assessed paper quality corresponds with the objective benchmarks commonly used by economists to evaluate research quality. However, an alternative empirical strategy is needed to analyze papers from unranked journals and AI-generated content, where an objective quality benchmark—such as RePEc's aggregate journal ranking—is unavailable.

To systematically compare papers from ranked journals, unranked journals, and AI-generated sources, we group the ranked journals into quartiles based on their position in the RePEc ranking. We then use OLS to estimate LLMs' assessment of paper quality across all groups using the following specification:

$$Y_i = \delta G_i + \theta_1 Unranked_i + \theta_2 AI_i + \beta_1 Length_i + \beta_2 Length_i^2 + \gamma_f + \varepsilon_i, \quad (3)$$

where $G_i$ is a categorical variable indicating the quartile of journal $j$ in which paper $i$ was published. The top quartile serves as the omitted reference category. For papers published in journals not ranked within RePEc, as well as AI-generated papers, $G_i$ is set to zero. $Unranked_i$ is a dummy variable with a value equal to 1 if paper $i$ was published in a journal not ranked within RePEc, and $AI_i$ is a dummy variable with a value equal to 1 if paper $i$ was generated by an AI model. As before, we compute heteroskedasticity-robust standard errors to account for potential variation in the residual variance across observations.

All analyses are conducted separately for each of the four LLMs (GPT, Claude, Gemma, and LLaMA). This approach avoids imposing any assumptions about shared evaluation patterns or error structure across AI models.

## 4. Results

Figure 1 displays nonparametric binscatter estimates of how each LLM assesses the quality of 1,000 unique academic papers based on journal ranking. The y-axis shows the average "Top 5 journal" rating assigned by each LLM to a given paper, and the x-axis represents the ranking of the journal in which the paper was published (lower values indicate higher-ranked journals). In Panel A, we utilize the ordinal ranking of journals within our selected sample (i.e., ranked 1 to 100). In Panel B, we employ each journal's actual RePEc ranking within the entire distribution, which extends beyond the top 100 and accurately reflects their position in the broader RePEc population of journals. Each subplot corresponds to one LLM: GPT, Claude, Gemma, and LLaMA.

For each model, the data are partitioned into 100 bins, with each bin containing 10 randomly selected papers. A linear polynomial is estimated within each bin, while a global cubic polynomial is employed for bias correction and to construct 95% confidence intervals (represented by vertical bars). The dashed line depicts the bias-corrected conditional expectation function, illustrating how the LLM's evaluation varies across the journal ranking distribution.

Across all models in both panels, there is a distinct downward-sloping relationship evident in the dashed line, indicating that papers published in higher-ranked journals–whether ordinally- or cardinally-measured—tend to receive higher "Top 5 journal" ratings from the LLMs. The shape and scale of the decline vary considerably across models. GPT and Gemma display the most pronounced nonlinear patterns, with sharper declines among top-ranked journals followed by flattening. Gemma also exhibits the widest range of ratings and the lowest overall level of perceived quality. In contrast, Claude and LLaMA show more gradual, approximately linear downward trends, with consistently higher baseline ratings—particularly for LLaMA, which appears least sensitive to journal rank.

Regarding the overall average rating, Claude and LLaMA tend to rate papers more generously than GPT and Gemma. The mean rating for all papers is 4.77 (*S.D.* = 0.69) for Claude and 5.72 (*S.D.* = 0.56) for LLaMA, while GPT and Gemma have mean ratings of 3.69 (S.D. = 0.58) and 3.74 (S.D. = 0.77), respectively. This pattern indicates a systematic variation in scoring generosity among the models.

To ensure that the patterns we observe are not sensitive to the choice of bin width, we also estimate the conditional expectation function using the IMSE-optimal number of bins, as determined by the automatic partitioning procedure in Cattaneo et al. (2025). Appendix C shows the results based on the IMSE-optimal number of bins, which ranges from 4 to 6 depending on the model. The downward relationship between journal ranking and LLM-assigned quality remains evident across all four models. Despite the reduced granularity, the overall pattern—particularly the steeper decline among higher-ranked journals—persists. This reinforces the robustness of the nonparametric trends observed in the fixed-bin specification.

Using the IMSE-optimal number of bins also enables formal inference on the shape of the conditional expectation function—such as tests for linearity or monotonicity—which are invalid with an arbitrarily fixed number of bins (e.g., 100). Across all models, we reject the null hypothesis of a zero-order polynomial—that is, no relationship between LLM-assigned quality scores and ordinal journal ranking—at the 1% significance level ($p < 0.001$). In contrast, we cannot reject the null hypothesis of a linear relationship for three of the four LLMs—GPT ($p = 0.208$), Claude ($p = 0.780$), and LLaMA ($p = 0.870$)—suggesting that a linear fit adequately captures the trend in those cases. However, for Gemma, we can reject linearity at the 10% level ($p = 0.084$). Still, we cannot reject the null hypothesis of a second-order polynomial ($p = 0.236$), indicating evidence of a quadratic relationship between journal ranking and Gemma-assigned quality.

We also formally test whether the conditional expectation function is weakly decreasing in journal rank—that is, whether LLM-assigned quality scores generally decline as journal prestige diminishes. For all four models, we do not reject the null hypothesis of monotonicity, with p-values significantly above 0.1. This is consistent with the observed negative relationship between RePEc journal rank and LLM-assigned quality scores, where more prestigious journals (indicated by a lower numerical rank) are consistently associated with higher perceived quality.

To gain insight into the potential mechanisms underlying how LLMs evaluate paper quality, Figure 2 replaces the "Top 5 journal" rating with LLM-generated scores for four specific dimensions: (i) originality, (ii) rigor, (iii) scope, and (iv) impact. Across all four dimensions, LLaMA seems largely unable to differentiate papers based on their originality, rigor, or scope, exhibiting minimal variation in scores throughout the journal ranking distribution. In contrast, Gemma demonstrates the greatest sensitivity to journal ranking, particularly in distinguishing papers based on the dimensions of originality, rigor, and impact. Given that originality and impact are among the strongest predictors of all LLMs' "Top 5 journal" ratings—see Appendix D for OLS regressions of each model's ratings on the four underlying domains—this pattern may help explain why Gemma emerges as the best-performing model in our analysis, while LLaMA appears least responsive to variation in journal ranking.

How do LLMs evaluate papers published in unranked journals or generated by AI? We address this question in Figure 3, which presents two distinct sets of results from OLS regressions outlined in Equation 3. These regressions compare LLM-assigned evaluations across ranked, unranked, and AI-generated papers, controlling for paper length and field fixed effects. The omitted reference group is the top-ranked journal quartile (Ranks 1–25). Panel A plots the estimated coefficients for each model's "Top 5 journal" rating, while Panel B shows coefficients from similar regressions where the dependent variable is the LLMs' "Written by AI" score, rated on a 0–10 scale. Together, these panels provide insight into how LLMs differentiate between traditional academic content and synthetic or lower-prestige writing—both in terms of perceived quality and likelihood of AI authorship.

The results in Panel A indicate that all four LLMs systematically assign lower "Top 5 journal" ratings to unranked papers, with estimated coefficients ranging from approximately –0.8 to –1.2 relative to the top-ranked journal group. This decline is significantly larger than the drop in ratings for papers published in lower-ranked journals (i.e., ranks 26–100), where estimated coefficients are generally around –0.5 or smaller in magnitude. For AI-generated papers, most LLMs rate them similarly to, or slightly below, those from the lowest-ranked quartile. GPT is the exception, assigning a moderately higher "Top 5 journal" rating to AI-generated papers (coefficient ≈ +0.5), suggesting it views them more favorably than human-written papers from unranked journals.

Panel B provides complementary insight into LLMs' perceived authorship detection. GPT stands out as the most consistent in identifying AI-generated papers, giving them significantly higher "AI-written" scores while assigning much lower scores to human-written papers, including those from unranked journals. This indicates that GPT is better at distinguishing synthetic content from less polished human writing. In contrast, Gemma assigns the highest "AI-written" scores to unranked papers—even higher than to actual AI-generated papers—suggesting that it may conflate lower-quality human writing with synthetic text. Claude and LLaMA exhibit intermediate patterns, also tending to over-attribute AI authorship to unranked papers, though not as strongly as Gemma.

One question of interest is whether certain LLMs are more capable of generating high-quality papers that are also more difficult to detect as AI-generated. To test this, we decompose the AI-generated papers by the specific LLM model used to create each one, distinguishing between three open-source models—Deepseek R1 (reasoning model), Gemma 3 27B, and LLaMA—and three closed-source models—GPTo1 (reasoning model), GPT-4o, and Claude 3.7 Sonnet (with extended reasoning capabilities). We then re-estimate Equation 3 and report the resulting coefficients in Appendix E. The reference group remains the same, which is the "1-25$^{th}$ ranked" category.

The results reveal clear differences in how LLMs assess both the quality and detectability of AI-generated academic papers, indicating that not all AI-written texts are evaluated equally—even within the same model group. For instance, papers generated by Claude receive the highest "Top 5 journal" ratings across nearly all evaluator models, with coefficients approaching or exceeding 1.0 in some instances, while those created by LLaMA receive significantly lower ratings, often close to zero or even negative. This suggests that papers produced by more advanced LLMs are not only more likely to be judged as high-quality but also more likely to exhibit variation in evaluation across different models.

Regarding AI detectability, Gemma consistently assigns the highest "Written by AI" ratings—often exceeding 3.0 on a 10-point scale—across all paper types, while the GPT, Claude, and LLaMA models cluster around much lower values (typically between 0.5 and 1.5). Notably, even for the same set of papers, evaluations can diverge sharply. For instance, Claude-generated papers are rated near 1.0 by GPT-4o in terms of AI detectability when compared to the top-ranked published papers but over 2.0 by Gemma, indicating a significant difference in what each model considers to be "AI-like" text.

One possible objection to our findings is that they may not generalize across different economic subfields. While our results remain robust after controlling for field fixed effects, it is still possible that the negative relationship between LLM evaluations and journal ranking is driven primarily by a small number of subfields. Figure 4 provides further reassurance that the relationship between journal rank and LLM-assessed quality is not driven by field-level heterogeneity. The figure plots the average "Top 5 journal" rating against journal rank separately for each field, along with field-specific linear fits. Despite some variation in baseline levels and slopes across fields, nearly all show a clear downward trend, indicating that LLMs associate higher-ranked journals with higher-quality papers within field. This pattern holds across all four models. Together with the inclusion of field fixed effects in all regressions, Figure 4's results support the conclusion that the observed

ranking–quality gradient reflects within-field evaluations, not differences in field composition.

We also conduct a robustness check on whether Gemma, the best-performing model in our analysis, adjusts its evaluation based on the amount of paper it reads. Figure 5 presents binscatter estimates of Gemma's "Top 5 journal" ratings against journal rank at varying levels of text exposure, from 0.1% to 100%. With minimal input (0.1% or 1% of the text), ratings remain relatively flat across the rankings, averaging around 3.0, which suggests limited signal extraction. As more text is provided, both the average rating and the slope of the relationship shift: at 50% and 100% of the text, the average rating rises to around 3.5–4.0, and a clear negative slope emerges. The findings suggest that Gemma's evaluations become more confident and discriminating with increased textual input, reinforcing the credibility of the observed ranking-quality gradient and reducing the likelihood that it arises by chance.

Finally, how biased are LLMs in their evaluation of single-blind submissions? To address this question, Table 1 presents OLS estimates of the effects of randomized author characteristics on each LLM's likelihood of assigning a "Top 5 Journal" rating. Robust standard errors—clustering at the paper ID level—are reported. Looking across columns, we find that—with the exception of Claude—GPT, Gemma, and LLaMA consistently rate papers attributed to top male and top female authors as significantly higher in quality than the same papers presented in the anonymized (blind) form. For example, GPT rates submissions by top male authors approximately 0.39 points (S.E.=0.042) higher than the same papers when submitted anonymously. This is a sizeable effect—larger, in fact, than the estimated difference between publishing in a journal ranked $51^{st}$–$75^{th}$ versus one ranked in the top $1^{st}$–$25^{th}$. There is also a noticeable penalty for submissions from top female authors—but not for randomly assigned female names. GPT, Gemma, and LLaMA all rate submissions by top female authors significantly lower than those by top male authors, with the differences statistically significant at the 1% level. The difference in ratings between random authors and the blind condition is statistically indistinguishable from zero.[4]

Additionally, all LLMs rate submissions from authors affiliated with MIT and Harvard significantly higher than the same papers presented anonymously. Meanwhile, three out of four LLMs—GPT, Gemma, and LLaMA—assign considerably higher ratings to submissions from authors affiliated with the London School of Economics (LSE). In contrast, Gemma rates submissions from all three non-US/UK universities—NTU, Chulalongkorn University, and the University of Cape Town—significantly lower than anonymized submissions. Claude, in particular, seems to heavily penalize submissions from Chulalongkorn University in Thailand.

It is also worth noting that across all LLMs, the coefficients associated with lower-ranked journals—relative to the top $1^{st}$-25th RePEc-ranked group—become increasingly negative as journal ranking declines. This pattern aligns with our earlier nonparametric analysis of journal quality. The results also suggest that earlier findings can be reliably replicated using a single evaluation per submission rather than three. This could substantially reduce operating costs associated with LLM use—particularly token-based API charges, compute time, and any human effort required to reconcile multiple outputs.

---

[4] Since we do not include lesser-known economists as a separate condition, we cannot definitively conclude whether LLMs favour economists—whether prominent or not—over random names, or whether they specifically favour top economists over random names. However, it is worth noting that in our earlier working paper, which included lesser-known economists as a condition, we rejected the null hypothesis and found that LLMs rated prominent economists significantly higher than lesser-known ones. See Pataranutaporn et al. (2025).

The relationship between a paper's length and LLMs' quality assessment follows an inverted U-shape, with the estimated maximum length averaging around 23,000 words. Compared to the applied micro-econometric journals, LLMs generally rate papers published in the Others—mostly econometrics journals—as higher, on average.

In summary, Table 1 provides strong evidence of AI bias in the single-blind review of economics papers, aligning with prior findings on bias in human peer review within the field of economics (Blank, 1991; Laband & Piette, 1994; Hengel, 2022; Huber et al., 2022). The findings also suggest that if editors wish to incorporate AI into the paper screening process, they should ensure that all author-identifying information is excluded when inputting the text into LLMs.

## 5. Theoretical Implications

In this section, we present a theoretical model that formalizes the role of editors in augmenting AI systems within the peer review process. Informed by this study's empirical insights, the model examines how integrating LLMs into editorial decision-making can help editors not only desk-reject low-quality or AI-generated papers more efficiently and accurately—while allowing genuinely high-quality submissions to move forward—but also reduce editorial bias toward prestige signals such as institutional affiliation, author gender, or well-known names. We begin by describing the existing system of the editorial decision-making process.

### 5.1. Pre-LLM Desk Decisions

In the traditional pre-LLM environment—where submission volumes can be overwhelming—editors often take considerable time to assess each paper before deciding whether to desk-reject it or send it out for review. They also spend significant time identifying suitable reviewers who are both qualified and willing to accept the invitation. Faced with these constraints, many editors are assumed to rely on relatively quick impressions of manuscripts, frequently supplementing those impressions with reputational cues such as the authors' institutional affiliations or personal prominence.

We assume the editor to form a quick heuristic score, $h_i^{pre}$. This score may be only weakly correlated with a manuscript's actual quality, and may explicitly reflect biases in favor of male authors and those with well-known names (Hengel, 2022; Huber et al., 2022). If we let $q_i \in \{0,1\}$, which denotes whether paper $i$ is genuinely high-quality ($q_i = 1$) or low-quality ($q_i = 0$), the editor's heuristic can be represented as:

$$h_i^{pre} = \gamma_q \tilde{q}_i + \gamma_a \psi(A_i) + v_i, \qquad (4)$$

where $\tilde{q}_i$ is the editor's noisy and oftentimes cursory impression of the paper's merit, $\gamma_q \geq 0$ indicates how strongly the editor weighs that impression, $\gamma_a \geq 0$ captures bias toward prominent authors, $\psi(A_i)$, and $v_i$ is editorial noise. A paper is desk-rejected if $h_i^{pre} < \delta_0$, which is the desk-rejection threshold that the editor applies when deciding whether to reject a paper outright without sending it for peer review. This heuristic approach can systematically penalize high-quality but lesser-known authors when $\gamma_a$ is large and $\tilde{q}_i$ is weakly correlated with the paper's true quality.

To make this more precise, assume $h_i^{pre}|q_i$ follows normal distribution with means $\mu_H^{pre}, \mu_L^{pre}$ and standard deviation $\Sigma_{pre}$. Define the probability of rejecting a good paper and accepting a bad one, then combine these into an expected loss function,

$$L^{pre}(\delta) = c_1 \pi \Phi\left(\frac{\delta - \mu_H^{pre}}{\Sigma_{pre}}\right) + c_2(1-\pi)[1 - \frac{\delta - \mu_L^{pre}}{\Sigma_{pre}})], \qquad (5)$$

where $\pi = Pr\ (q_i = 1)$ is the fraction of high-quality papers, $c_1 > 0$ is the cost of wrongly rejecting a good submission, and $c_2 > 0$ is the cost of wrongly accepting a bad one. Minimizing $L^{pre}(\delta)$ with respect to the threshold $\delta$ yields a first-order condition by setting $\frac{dL^{pre}}{d\delta} = 0$, which ultimately pins down an optimal $\delta^{pre*}$. However, because $\mu_H^{pre} - \mu_L^{pre}$ reflects the editor's noisy heuristic evaluations—based on superficial reading and prestige-related cues—the separation between good and bad papers in the score distribution may be limited, thereby complicating the task of identifying an optimal editorial threshold. As a result, the signal quality in the pre-LLM environment—where efficiency is a primary concern—is often compromised in ways that are not only suboptimal but also inequitable, as high-quality submissions from less prestigious or underrepresented authors are more likely to be wrongly rejected.

### 5.2. Post-LLM Desk Decisions

In the AI-augmented review process, the editor receives a composite LLM score

$$s_i^{LLM} = \omega_q\left(\alpha_q q_i + \eta_{i,Q}\right) + \omega_f\left(\alpha_f \delta(paper\ genuine) + \zeta_{i,F}\right), \qquad (6)$$

where $\alpha_q q_i + \eta_{i,Q}$ captures refined textual-quality analysis that the editor obtains from an LLM specialized in identifying strong research papers–such as the existing versions of Gemma or Claude. These models exhibit a steep gradient between textual signals and recognized benchmarks of scholarly merit, implying a relatively large $\alpha_q q_i$. The notation $\alpha_f \delta(paper\ genuine) + \zeta_{i,F}$ represents an authenticity score from a second LLM (e.g., a combined usage of Gemma and GPT) that specializes in spotting suspicious or AI-generated text. Both sub-models see only anonymized text, so they are immune to author-based biases. We then define the desk-rejection rule as

$$d_i^{post} = 1\{s_i^{LLM} + \gamma_a \psi(A_i) < \delta\}. \qquad (7)$$

A paper with strong textual qualities but no prestige signals can still survive, provided $s_i^{LLM}$ is sufficiently high to overcome $\gamma_a \psi(A_i)$. Moreover, a polished by fraudulent paper may yield a low authenticity score, pushing $s_i^{LLM}$ below the desk threshold. If the LLM signals under good and bad conditions have means $\mu_H^{post}, \mu_L^{post}$ and standard deviation $\Sigma_{post}$, then the expected loss for post-LLM editorial decision-making can be written as:

$$L^{post}(\delta) = c_1 \pi \Phi\left(\frac{\delta - \mu_H^{post}}{\Sigma_{post}}\right) + c_2(1-\pi)[1 - \frac{\delta - \mu_L^{post}}{\Sigma_{post}})]. \qquad (8)$$

Taking the derivative with respect to $\delta$ and setting it to zero identifies $\delta^{post*}$, the optimal LLM-based desk threshold.

### 5.3. Derivative-based Comparison and Efficiency

Let us define the signal separation measures of pre- and post-LLM as:

$$\Delta^{pre} = \frac{\mu_H^{pre} - \mu_L^{pre}}{\Sigma_{pre}}, \Delta^{post} = \frac{\mu_H^{post} - \mu_L^{post}}{\Sigma_{post}}. \qquad (10)$$

If the LLM-based signal more effectively separates high- and low-quality papers compared to the editor's heuristic, then $\Delta^{post} > \Delta^{pre}$. Since the expected loss functions $L^{pre}$ and $L^{post}$ both follow the same normal form but with different mean and variance parameters, having a larger separation $\Delta$ translates into lower classification error at the threshold that minimizes each function's derivative, i.e., $L^{post*} < L^{pre*}$. In other words, LLMs can more reliably recognize high-quality papers (thereby limiting false rejections) and more effectively flag poor or fabricated papers (reducing false acceptances). This reduces overall screening loss, which can arise as a trade-off when attempting to shorten editorial turnaround times. In addition, the availability of an LLM-based continuous quality score enables editors to set their screening thresholds, optimizing the trade-off between quality control and resource constraints. A sufficiently discriminative signal allows for more effective desk rejections, thereby reducing the number of papers sent out for peer review without increasing overall screening loss.

## 6. Discussion and Conclusion

This study presented one of the first large-scale experimental evaluations of large language models (LLMs) as potential tools for enhancing efficiency without compromising accuracy in the academic peer review process in economics. Using a combination of nonparametric binscatter and linear regressions on over 29,000 evaluations across 1,220 papers—including published, unranked, and AI-generated submissions—we found that LLMs could reliably distinguish between higher- and lower-quality research based solely on textual content, even without access to author or journal metadata.

More specifically, we showed that LLMs consistently assigned higher quality scores to papers published in top-ranked RePEc journals and lower scores to those from lower-ranked or unranked outlets. This trend was evident across economic subfields and remained robust under various controls, including paper length and field fixed effects. Among the models tested, Claude and Gemma displayed the greatest sensitivity to journal rankings, producing the most pronounced quality gradients. Gemma was particularly effective at penalizing low-quality and AI-generated submissions, whereas GPT—although less responsive to journal rank—outperformed the others in detecting AI authorship. LLaMA, by contrast, showed minimal responsiveness to variation in journal rank, limiting its utility for quality-based screening.

Our findings also highlight that evaluations of AI-generated research depend not only on the model that produced the content but also on how different evaluator models interpret it. This makes model choice crucial—not just for generating text but for downstream tasks like peer review, AI-authorship detection, and policy decisions. The variation across LLM evaluators also cautions against placing too much weight on automated assessments when judging the quality or authenticity of AI-generated work.

We also conducted an additional experiment involving 8,910 evaluations of 330 papers spanning the journal ranking distribution to examine whether LLMs exhibited bias against female authors, individuals from lesser-known institutions, or less prominent authors in single-blind submissions relative to fully anonymized ones. By systematically varying author characteristics across papers, we provided clear evidence of AI favoritism toward submissions from top male authors and those affiliated with elite institutions. These findings suggest that editors considering AI-assisted pre-screening should take care to remove all author-identifying information before inputting manuscripts into LLMs to minimize the risk of bias.

Our results are not without valid objections—one of which is that not all papers published in top-ranked journals are necessarily of higher quality than those in lower-ranked outlets. This raises the concern that RePEc's journal rankings may not serve as a reliable benchmark for assessing the quality of individual papers. This critique is valid, as prior research has shown that many papers published in top journals receive fewer citations than those appearing in mid-tier outlets (e.g., Oswald, 2007). Nonetheless, while such exceptions undoubtedly exist, our large-scale and generalizable study demonstrates that, on average, the evaluated LLMs rated the textual quality of papers from top-ranked journals higher than that of papers from lower-ranked outlets. That said, it remains debatable whether LLMs should have assigned substantially higher ratings to papers published in top-five journals compared to those in journals ranked only slightly lower—especially considering the astronomical weight many economists place on publishing in top-five outlets.

Another objection is that none of the LLMs recommended rejection—a score of 1 or 2—for more than 90% of the published papers in RePEc-ranked journals, even when asked to rate them based on their likelihood of acceptance at a top-five journal. This could suggest that the LLMs regarded most published papers as potentially revisable, or alternatively, that they are systematically biased in a positive direction—assigning a uniformly high baseline score to all papers, while still differentiating between higher- and lower-quality work through the slope of their evaluations. As a result, editors should apply a calibrated acceptance threshold that combines LLM evaluations with expert editorial judgment to account for disciplinary practices, methodological diversity, and innovative approaches, rather than relying solely on the categorical rating scale. Future research should explore different methodological approaches to establish optimal thresholds that balance efficiency gains with minimizing false negatives.

One limitation of our findings is that we cannot directly compare the effect sizes of AI biases with those of human reviewers. Although prior research provides insights into the magnitude of human bias favoring prominent authors (e.g., Huber et al., 2022), our study is not directly comparable, as effect sizes likely depend on context. For instance, our analysis was based on already published papers drawn from 110 journals, whereas the paper examined by Huber et al. (2022) was a working paper that had not yet undergone peer review and might still require substantial revision before publication. This difference may partly explain the larger disparity in rejection rates between prominent and lesser-known authors observed in their study. Moreover, conducting a comparable large-scale experiment on human reviewers to elicit their revealed preferences in the same way we did with AI would be highly challenging.

Another limitation of our findings on AI bias lies in the experimental design. While we systematically varied author characteristics such as institutional affiliation, reputation, and gender, we did not consider other potentially significant factors, such as race—which may be inferred from author names (Bertrand & Mullainathan, 2004)—or the geographic location of an author's institution. Biases against certain ethnicities, cultural identities, or regions may similarly influence both AI and human evaluations, but these dimensions remain unexplored in our study. This omission is noteworthy given the increasing diversity of contributors to global academic research and the growing emphasis on regional representation in economics. Future research should address this gap by examining whether and how racial and regional biases affect evaluations by LLMs. Nonetheless, while further exploration of AI bias is undoubtedly important, as we have argued, editors can already take a straightforward step to mitigate such bias—by ensuring that all author-identifying information is excluded before submitting manuscripts to LLMs.

Furthermore, a consideration arises regarding the potential for adversarial manipulation of LLM evaluations. The integrity of the algorithmic assessment system may be compromised through the deliberate prompt injection designed to circumvent the editor's parameter of language models (Perez et al., 2022; Wei et al., 2023). Such adversarial inputs necessitate robust preprocessing protocols to identify and eliminate potentially subversive content prior to LLM evaluation, thereby preserving evaluation validity. Our findings suggest that journal editors should implement a procedural safeguard—specifically, the systematic redaction of author-identifying information and malicious prompt injection before manuscript submission to LLMS for evaluation.

Despite these limitations, our study provides compelling evidence for integrating AI into the peer review process. As demonstrated in the theoretical implications section, the expected loss regarding both accuracy and equity is lower under an AI-assisted editorial regime—mainly when LLMs operate on anonymized submissions—compared to the pre-LLM environment, where decisions rely more heavily on potentially biased heuristics. This approach also has significant potential to enhance efficiency and reduce turnaround times for desk decisions. Furthermore, as long as editors can calibrate their screening thresholds, the number of manuscripts requiring full review should decrease without increasing the overall screening loss.

Finally, while our findings indicate that LLMs can effectively distinguish between high- and low-quality papers, we do not believe they can—or should—entirely replace human reviewers. Human judgment remains crucial, especially in assessing methodological rigor, contextual relevance, new developments, future implications, and identifying fraudulent data—areas that LLMs, trained on existing internet content, are not yet equipped to fully grasp.

More generally, the conclusions drawn from this research extend beyond publishing in economics. As journals increasingly face pressures to accelerate the peer review process without compromising quality, the idea of integrating AI systems in the peer review process offers a path forward. However, achieving this will require a concerted commitment from journals and the academic community to uphold transparency, fairness, and rigorous evaluation of LLM’s impact on equity, efficiency, and ethical integrity in scholarly publishing.

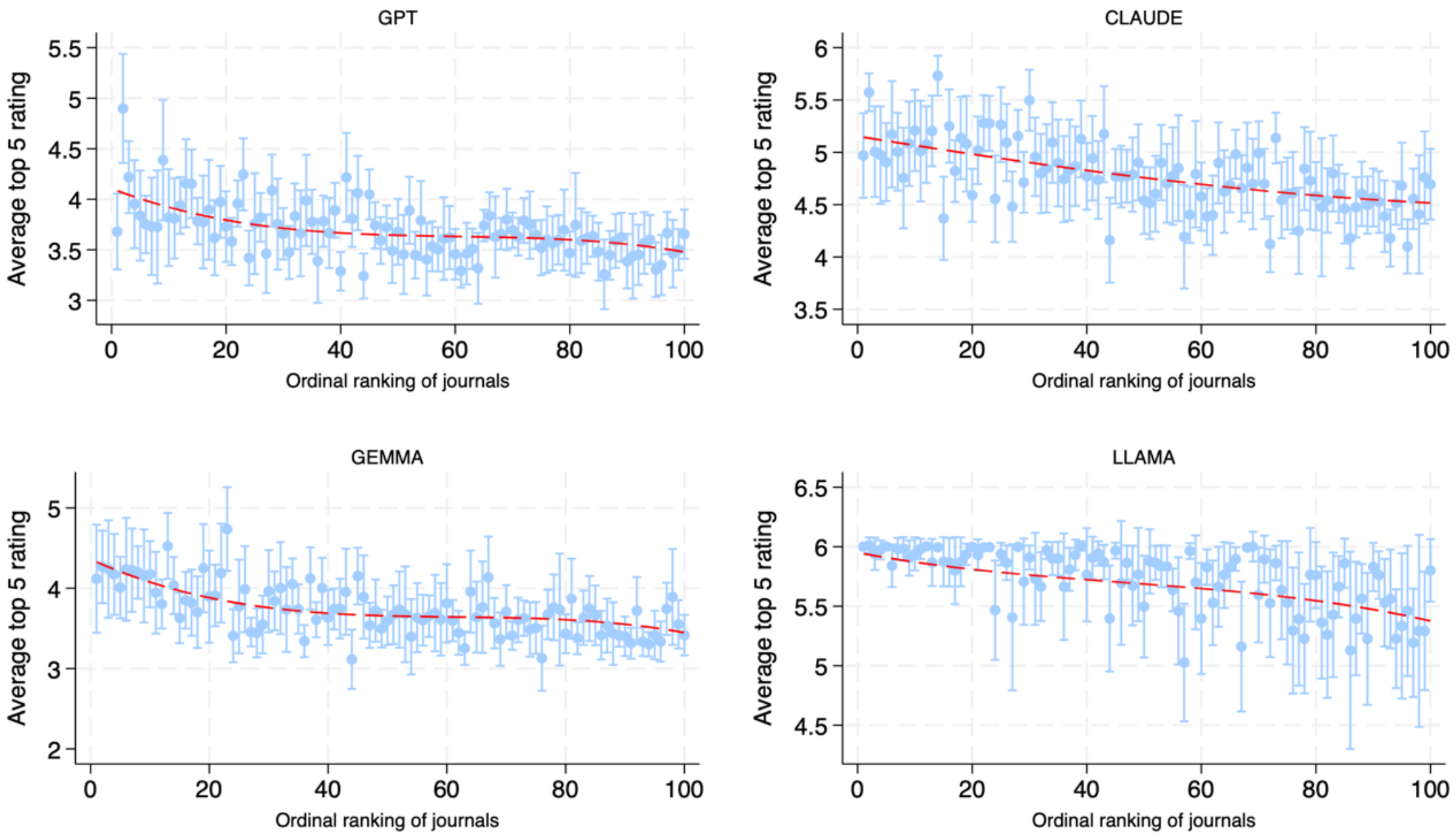

**A:** Top 5 ratings and ordinal ranking of journals on RePEc (February 2025)

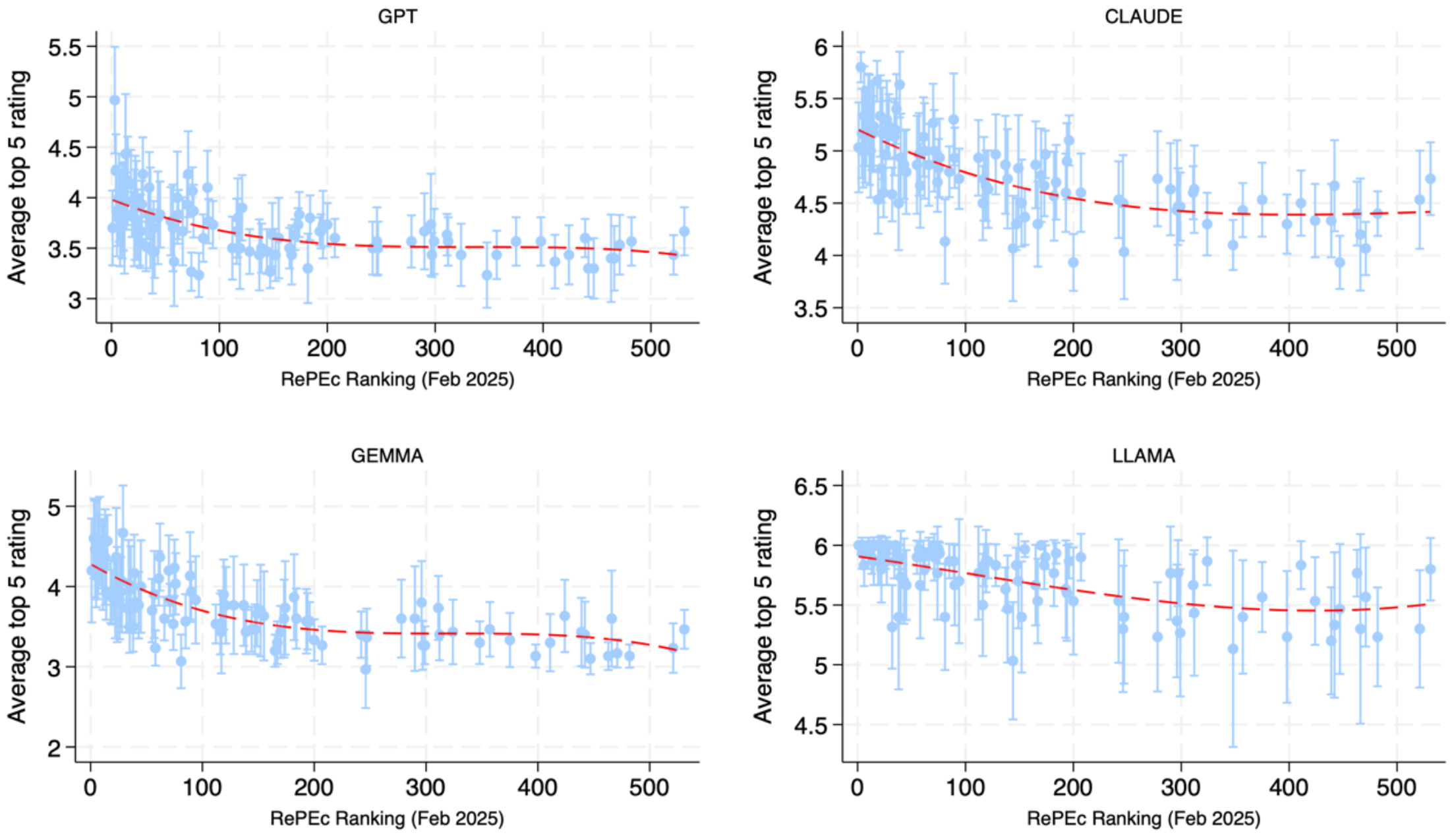

**B*:*** Top 5 ratings and actual ranking of journals on RePEc (February 2025)

**Fig 1**: Plots of binscatter least square estimates of LLMs' "Top 5 journal" evaluation of 100 economics journals—based on 10 recently published papers per journal—against RePEc's aggregate journal rankings as of February 2025. All estimates are adjusted for paper length, length squared, and journal field dummies (Microeconomics, Macroeconomics, Applied Micro-econometrics, General, and Finance). Panel A uses the ordinal ranking of the journals included in the analysis (e.g., *Econometrica* is assigned rank 1, *American Economic Review* rank 3). Panel B uses RePEc's actual ranking, where *Econometrica* remains 1st, but *American Economic Review* is ranked 4th because the *Journal of Economic Literature* (ranked 3rd) is not in the sample. Confidence intervals are constructed using bias correction based on a global cubic polynomial fit, and the 95% pointwise confidence intervals are based on that same global fit. The dashed line represents the bias-corrected conditional expectation function.

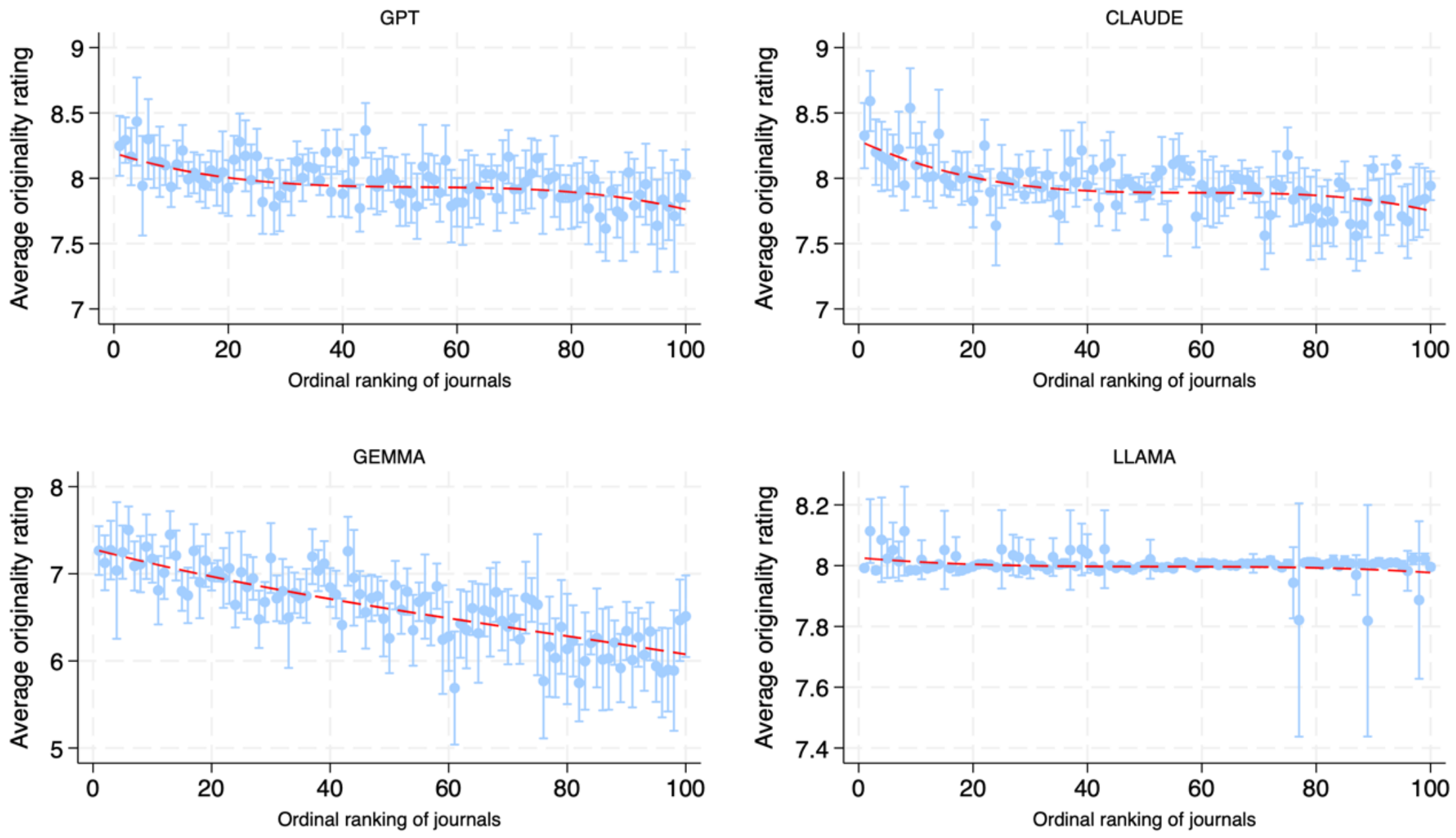

**A:** Originality rating

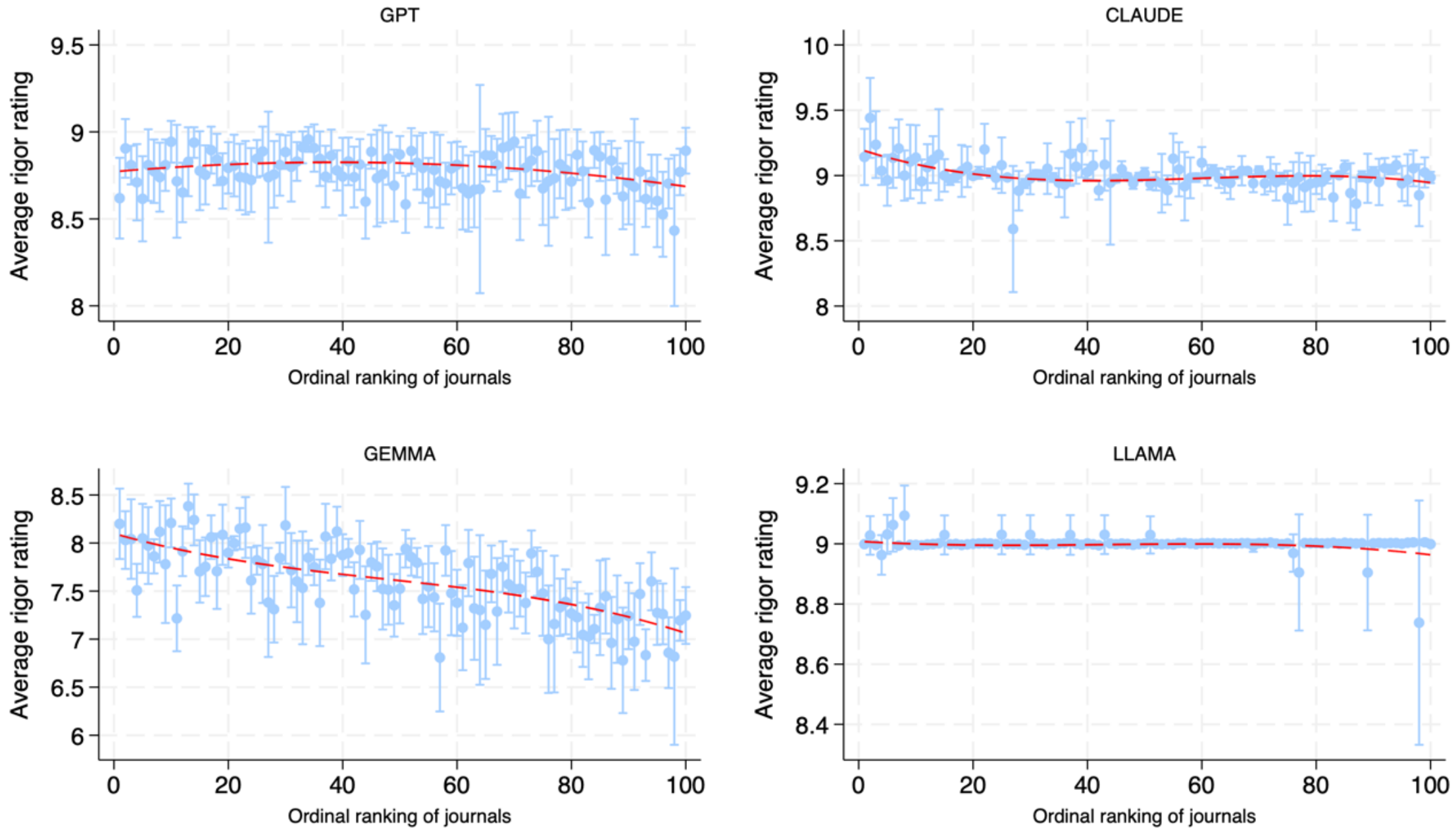

**B:** Rigorous rating

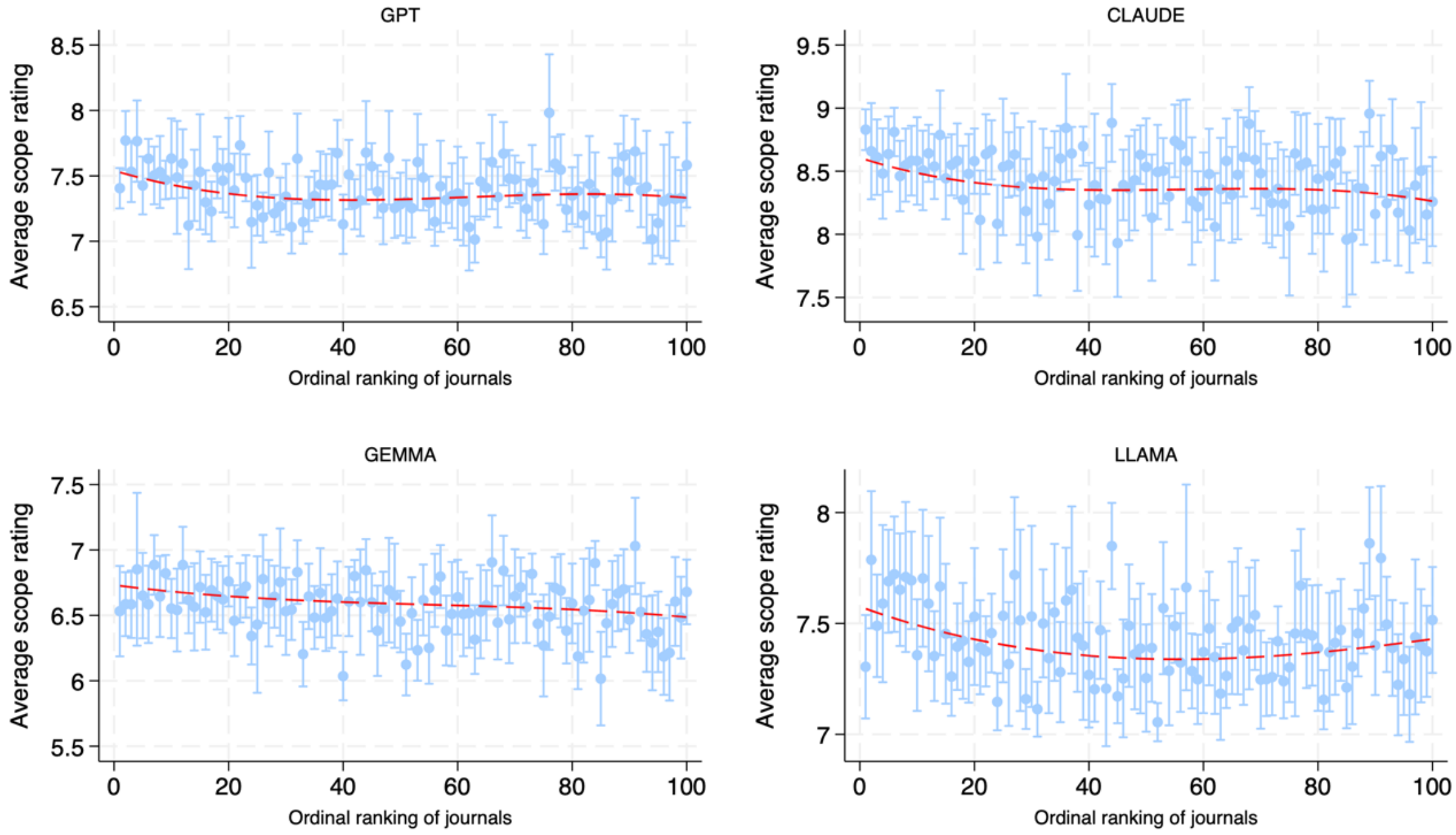

**C:** Scope rating

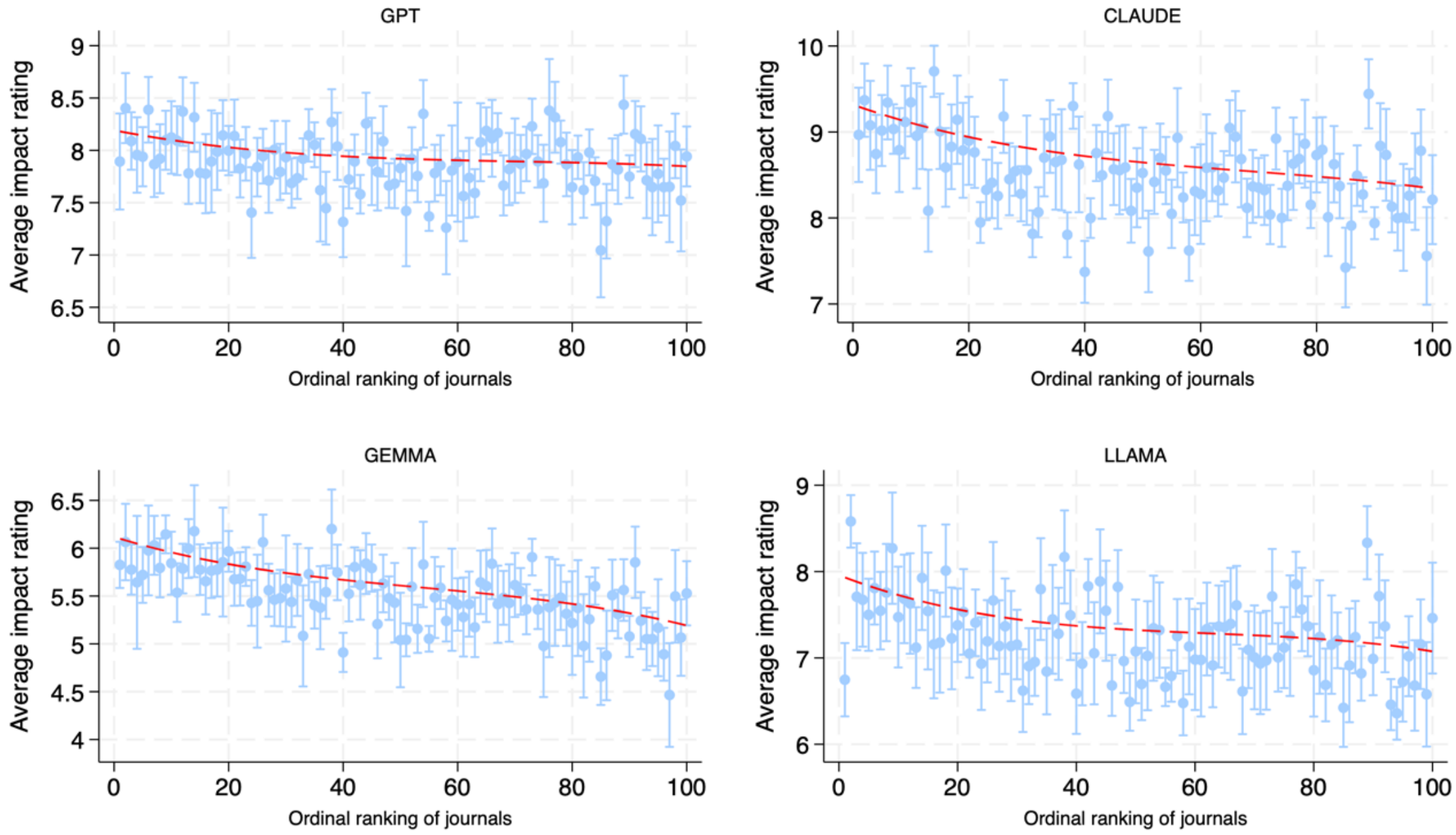

**D:** Impact rating

**Fig 2:** Plots of binscatter least square estimates of LLMs' evaluation on each paper's originality, rigor, scope, and impact against RePEc journal ranking. All ratings are on a scale of 1-10. The same note applies as in Figure 1.

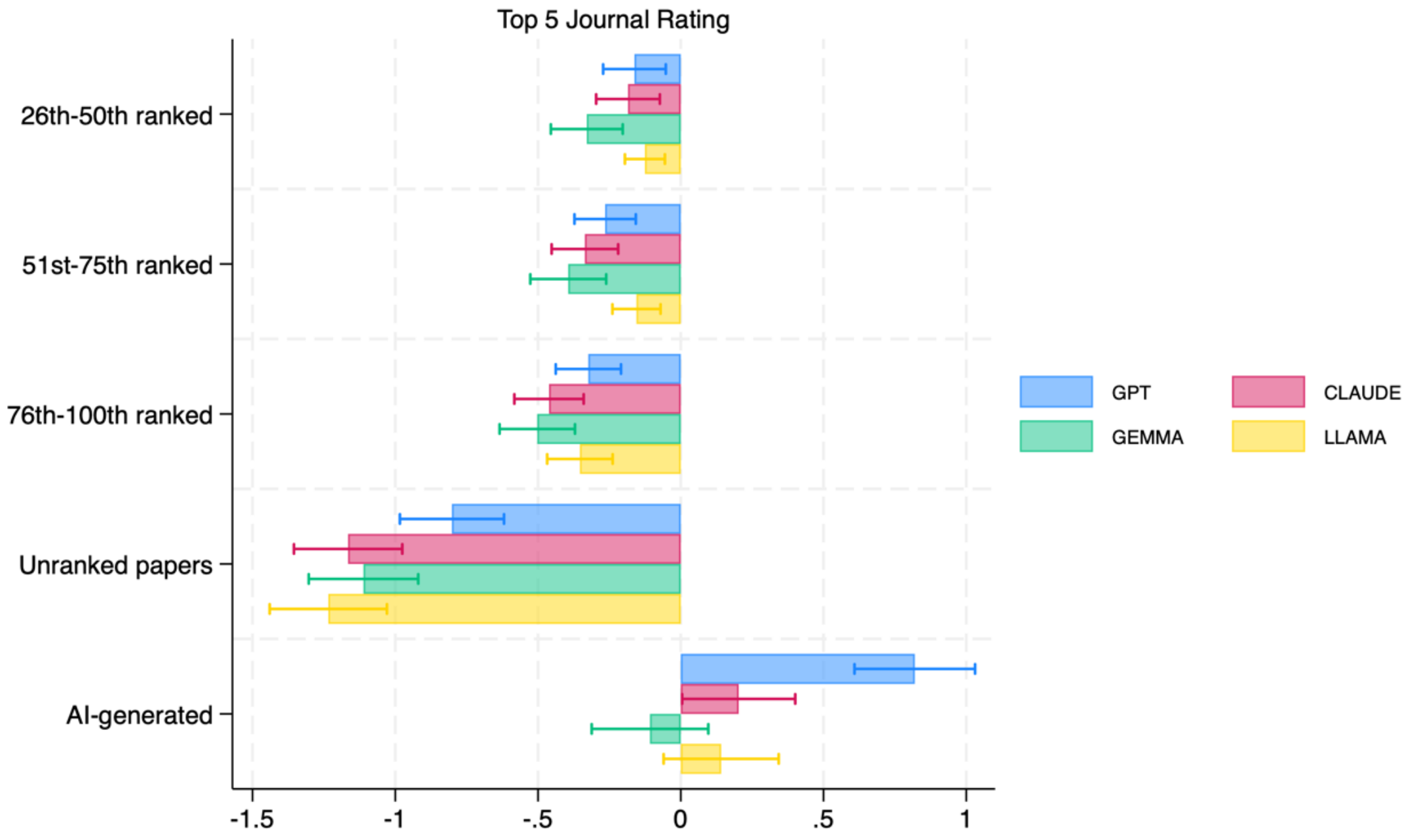

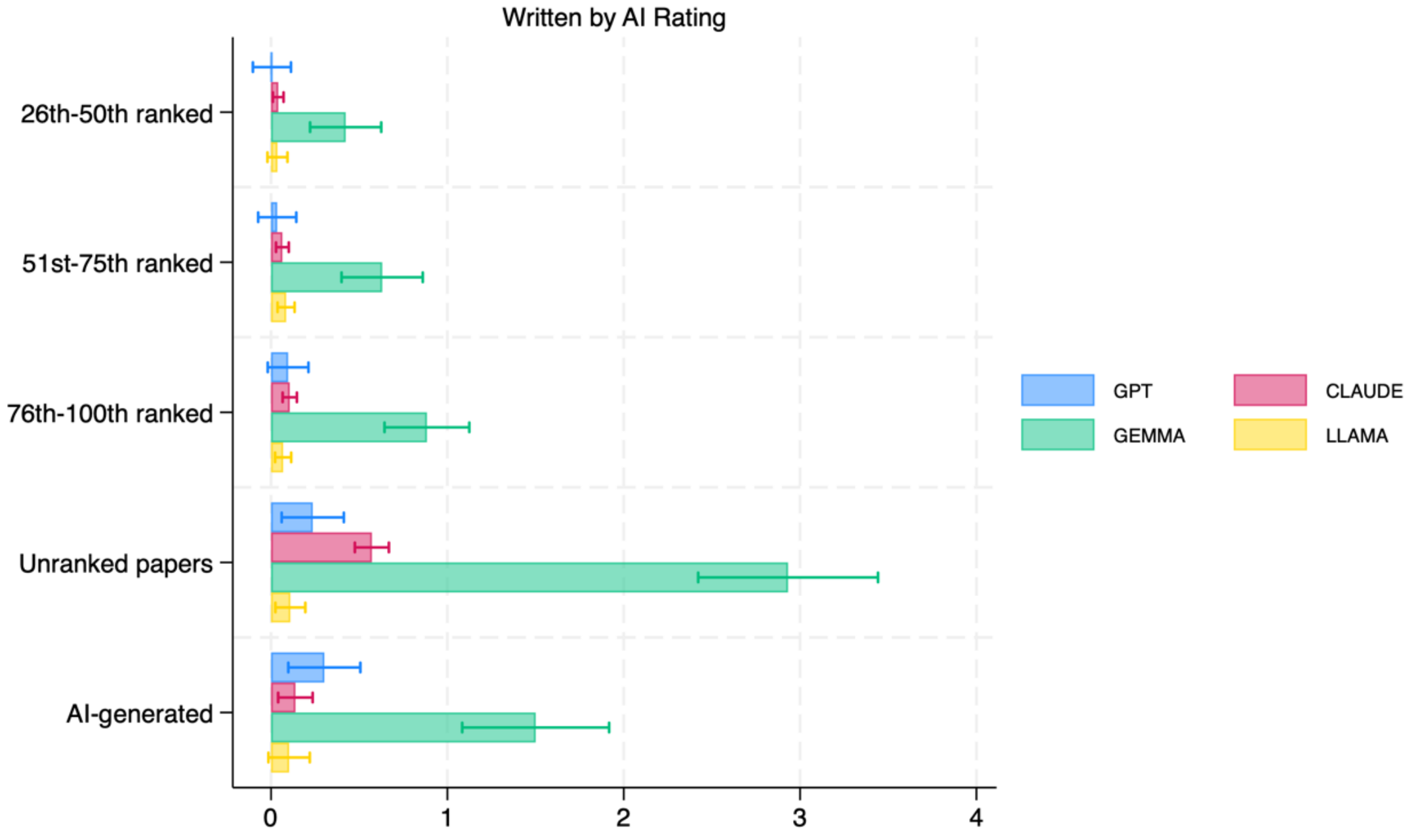

**Fig. 3.** Coefficient plots for each LLM model from regressions on the “Top 5 journal” rating (1–6 scale) and the “Written by AI” rating (1–10 scale). The reference category is journals ranked 1st–25th. All regressions include controls for paper length, paper length squared, and field fixed effects. 95% confidence intervals are shown.

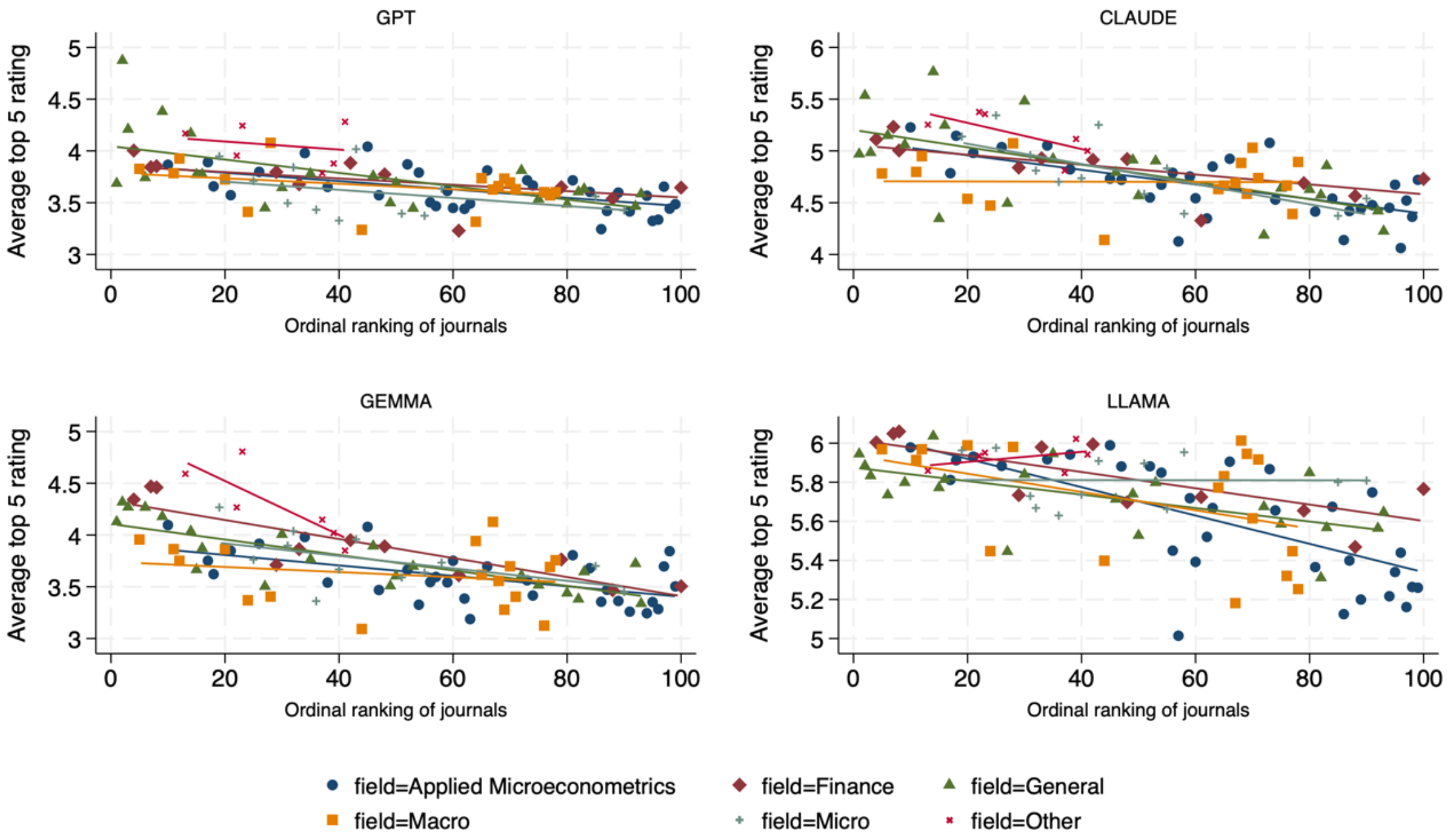

**Fig 4.** Plots of binscatter least square estimates of LLMs' "Top 5 journal" evaluation of 100 economics journals—based on 10 recently published papers per journal—against RePEc's aggregate journal rankings as of February 2025 by economic subfields. The best-fit lines are constructed using bias correction based on a global linear polynomial fit.

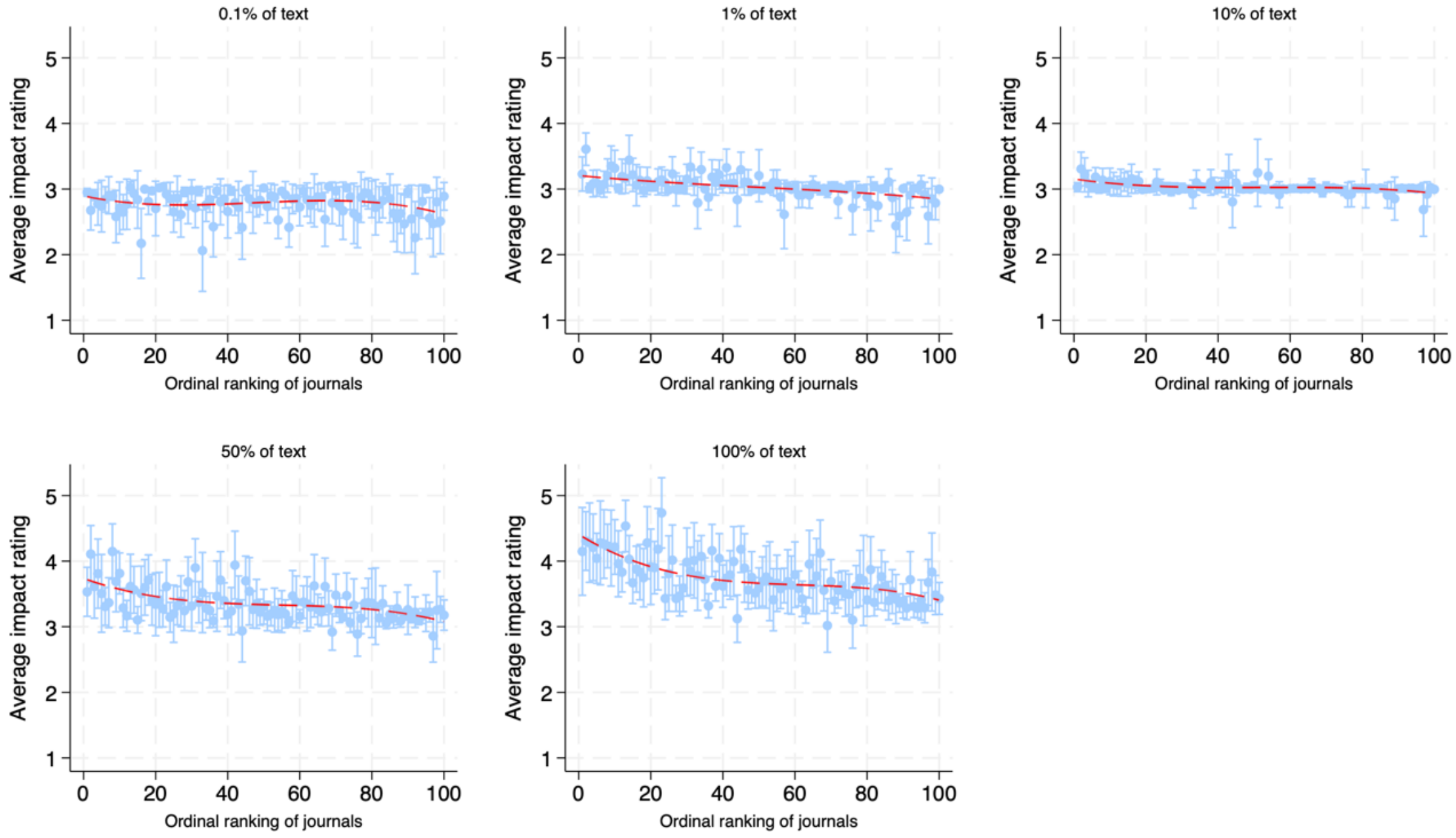

**Fig 5. Binscatter Plots of Average "Top 5 Journal" Ratings by Journal Rank Across Varying Text Exposure Levels (Gemma).** Each panel shows a binscatter plot of average impact ratings assigned by Gemma against the ordinal ranking of journals, using increasing proportions of paper text (0.1%, 1%, 10%, 50%, and 100%). The same note applies as in Figure 1.

## Table 1: Predicting LLMs' Top 5 Journal Ratings by Randomized Author's Characteristics: Ordinary Least Squares

| Variables | (1)<br>GPT<br>Top 5<br>rating | (2)<br>Claude<br>Top 5<br>rating | (3)<br>Gemma<br>Top 5<br>rating | (4)<br>LLaMA<br>Top 5<br>rating |
|---|---|---|---|---|
| **Author's name and gender** | | | | |
| Top male authors | 0.393*** | 0.026 | 0.401*** | 0.230*** |
| | (0.042) | (0.039) | (0.030) | (0.031) |
| Top female authors | 0.289*** | -.039 | 0.347*** | 0.181*** |
| | (0.040) | (0.036) | (0.028) | (0.030) |
| Random male authors | 0.085** | -0.053 | 0.014 | 0.029 |
| | (0.037) | (0.035) | (0.022) | (0.025) |
| Random female authors | 0.103*** | -0.081** | 0.029 | 0.040 |
| | (0.037) | (0.035) | (0.022) | (0.026) |
| **Author's institution** | | | | |
| MIT | 0.221*** | 0.115*** | 0.118*** | 0.109*** |
| | (0.049) | (0.042) | (0.032) | (0.031) |
| Harvard | 0.242*** | 0.082* | 0.118*** | 0.179*** |
| | (0.049) | (0.045) | (0.030) | (0.032) |
| LSE | 0.145*** | -.012 | 0.055* | 0.070** |
| | (0.048) | (0.045) | (0.030) | (0.035) |
| University of Cape Town | 0.052 | -0.061 | -0.045* | 0.030 |
| | (0.048) | (0.043) | (0.025) | (0.031) |
| Nanyang Technological University | 0.033 | 0.033 | -0.045* | 0.070** |
| | (0.049) | (0.042) | (0.025) | (0.034) |
| Chulalongkorn University | 0.030 | -0.148*** | -0.082*** | -0.015 |
| | (0.047) | (0.043) | (0.028) | (0.031) |
| **RePEc ordinal ranking** | | | | |
| $26^{th}$-$50^{th}$ ranked | -0.295*** | -0.417*** | -0.396*** | -0.131*** |
| | (0.077) | (0.084) | (0.105) | (0.049) |
| $51^{st}$-$75^{th}$ ranked | -0.302*** | -0.351*** | -0.494*** | -0.086 |
| | (0.079) | (0.091) | (0.110) | (0.070) |
| $76^{th}$-$100^{th}$ ranked | -0.465*** | -0.609*** | -0.584*** | -0.268*** |
| | (0.083) | (0.097) | (0.115) | (0.086) |
| Unranked papers | -1.185*** | -1.443*** | -1.445*** | -1.009*** |
| | (0.136) | (0.156) | (0.203) | (0.160) |
| **Paper's length** | | | | |
| Paper's length/100 | 0.009*** | 0.018*** | 0.012*** | 0.011*** |
| | (0.002) | (0.003) | (0.003) | (0.003) |
| (Paper's length/100-squared)/100 | -.002*** | -.004*** | -.002*** | -.003*** |
| | (0.001) | (0.001) | (0.001) | (0.001) |
| **Journal field** | | | | |
| Finance | -0.040 | -0.030 | 0.161 | 0.060 |
| | (0.084) | (0.133) | (0.155) | (0.094) |
| General | 0.134* | 0.058 | 0.196* | 0.072 |
| | (0.078) | (0.084) | (0.106) | (0.067) |
| Macroeconomics | 0.015 | -.006 | 0.032 | -0.038 |
| | (0.068) | (0.089) | (0.110) | (0.095) |
| Microeconomics | -0.088 | 0.180* | 0.016 | 0.189*** |
| | (0.074) | (0.099) | (0.105) | (0.064) |
| Others | 0.374*** | 0.238* | 0.440** | 0.152** |
| | (0.107) | (0.132) | (0.184) | (0.063) |
| Intercept | 3.210*** | 3.659*** | 2.893*** | 4.842*** |
| | (0.203) | (0.236) | (0.264) | (0.235) |
| Observations | 8910 | 8910 | 8910 | 8910 |
| $R^2$ | 0.165 | 0.323 | 0.344 | 0.246 |

**Notes:** Robust standard errors are reported in parentheses. *** $p < 0.01$, ** $p < 0.05$, * $p < 0.1$. The reference category for RePEc ordinal ranking is "1st–25th ranked," for journal field is "Applied micro-econometrics," and for the author's name, gender, and institution, it is "blind submission."

## Appendix A: List of Journals and their RePEc Ranking (February 2025)

| Journal | Average paper length | Ordinal rank | Actual rank on RePEc | Journal subfield |
|---|---|---|---|---|
| Econometrica | 15093.7 | 1 | 1 | General |
| Quarterly Journal of Economics | 20459.7 | 2 | 3 | General |
| American Economic Review | 18046.3 | 3 | 4 | General |
| Journal of Financial Economics | 18151.7 | 4 | 7 | Finance |
| American Economic Journal-Macro | 15935.9 | 5 | 5 | Macro |
| Journal of Political Economy | 18134.4 | 6 | 6 | General |
| Journal of Finance | 20742.9 | 7 | 8 | Finance |
| Review of Financial Studies | 21399.9 | 8 | 12 | Finance |
| Review of Economic Studies | 17689.6 | 9 | 13 | General |
| American Economic Journal-Applied Econ | 18035.3 | 10 | 10 | Applied micro |
| Journal of Economic Growth | 20008.4 | 11 | 11 | Macro |
| Journal of Monetary Economics | 16212.6 | 12 | 14 | Macro |
| Journal of Econometrics | 14633.8 | 13 | 15 | Others |
| Review of Econ and Stats | 11802.9 | 14 | 18 | General |
| Journal of Euro Econ Assoc | 18322.3 | 15 | 19 | General |
| Economic Journal | 15937.6 | 16 | 21 | General |
| Journal of Labor Economics | 13977.8 | 17 | 22 | Applied micro |
| American Economic Journal-Econ Policy | 18067.5 | 18 | 27 | Applied micro |
| RAND Journal of Economics | 15887.5 | 19 | 23 | Micro |
| Journal of International Economics | 13652.6 | 20 | 24 | Macro |
| Journal of Public Economics | 12436.5 | 21 | 26 | Applied micro |
| Journal of Bus and Econ Stats | 11390.8 | 22 | 28 | Others |
| Journal of Applied Econometrics | 12407.4 | 23 | 29 | Others |
| Journal of Development Economics | 14729.4 | 24 | 32 | Macro |
| Journal of Economic Theory | 17410.5 | 25 | 36 | Micro |
| Journal of Human Resources | 14344.1 | 26 | 34 | Applied micro |
| European Econ Review | 15458.5 | 27 | 38 | General |
| Journal of Money, Credit & Banking | 14634.4 | 28 | 35 | Macro |
| Journal of Banking and Finance | 17862.0 | 29 | 43 | Finance |
| International Econ Review | 17956.7 | 30 | 39 | General |
| Experimental Econ | 12874.5 | 31 | 40 | Micro |
| Journal of Econ Dynamic and Control | 13711.8 | 32 | 45 | Micro |
| Journal of Financial and Quant Analysis | 16641.3 | 33 | 57 | Finance |
| Journal of Urban Econ | 14531.3 | 34 | 61 | Applied micro |
| Oxford Bulletin of Econ and Stat | 13384.7 | 35 | 55 | General |
| Journal of Risk and Uncertainty | 12030.3 | 36 | 58 | Micro |
| Econometric Theory | 18518.2 | 37 | 62 | Others |
| Journal of Health Econ | 13543.7 | 38 | 66 | Applied micro |
| Quantitative Econ | 17465.2 | 39 | 70 | Others |
| Games and Econ Behavior | 12045.8 | 40 | 74 | Micro |
| Econometric Journal | 16010.5 | 41 | 71 | Others |
| Journal of Corporate Finance | 19386.2 | 42 | 76 | Finance |
| American Economic Journal-Microecon | 17037.7 | 43 | 89 | Micro |
| World Development | 12696.3 | 44 | 81 | Macro |
| Labour Economics | 15120.6 | 45 | 75 | Applied micro |
| Scandinavian J of Econ | 12563.5 | 46 | 94 | General |
| Journal of Econ Behav & Org | 14079.2 | 47 | 85 | Applied micro |
| Journal of Empirical Finance | 17938.0 | 48 | 90 | Finance |
| Canadian J of Econ | 14334.3 | 49 | 112 | General |
| Econ Inquiry | 12653.1 | 50 | 116 | General |
| Theoretical Economics | 18612.0 | 51 | 119 | Micro |
| American Journal of Agri Econ | 14151.9 | 52 | 121 | Applied micro |
| Economica | 15560.1 | 53 | 128 | General |
| Regional Science and Urban Econ | 14428.3 | 54 | 117 | Applied micro |
| Economic Theory | 16263.7 | 55 | 137 | Micro |
| Public Choice | 10678.5 | 56 | 139 | Applied micro |
| Regional Studies | 10355.7 | 57 | 144 | Applied micro |
| Journal of Mathematical Econ | 12719.5 | 58 | 155 | Micro |
| Econ of Education Review | 14643.4 | 59 | 149 | Applied micro |
| Journal of Econ Inequality | 11883.7 | 60 | 152 | Applied micro |
| International Finance | 11355.3 | 61 | 147 | Finance |
| Journal of Econ Psych | 11178.5 | 62 | 167 | Applied micro |
| Empirical Econ | 12755.5 | 63 | 165 | Applied micro |
| China Economic Review | 12002.5 | 64 | 182 | Macro |
| Economic Modelling | 12907.7 | 65 | 173 | Macro |
| Review of Income and Wealth | 13062.1 | 66 | 174 | Applied micro |
| Journal of Development Studies | 11196.7 | 67 | 32 | Macro |
| Review of International Econ | 11226.6 | 68 | 170 | Macro |
| Macroeconomic Dynamics | 26851.3 | 69 | 184 | Macro |
| The World Economy | 11182.8 | 70 | 194 | Macro |
| Journal of Macroeconomics | 11122.0 | 71 | 207 | Macro |
| Applied Economics | 9000.2 | 72 | 200 | General |
| Health Economics | 12535.3 | 73 | 196 | Applied micro |
| Journal of Regional Science | 15309.7 | 74 | 193 | Applied micro |
| Kyklos | 11863.9 | 75 | 242 | General |
| Journal of African Economics | 10937.9 | 76 | 246 | Macro |
| Journal of Policy Modeling | 8499.8 | 77 | 247 | Macro |
| Review of Development Economics | 10850.4 | 78 | 278 | Macro |
| Quarterly Review of Econ and Finance | 11621.1 | 79 | 290 | Finance |
| Manchester School | 10780.1 | 80 | 297 | General |
| Education Economics | 12367.9 | 81 | 296 | Applied micro |
| Empirica | 11641.4 | 82 | 299 | General |
| Scottish Journal of Pol Econ | 9840.1 | 83 | 312 | General |
| Papers in Regional Science | 13280.8 | 84 | 311 | Applied micro |
| Theory and Decision | 9826.6 | 85 | 324 | Micro |
| Journal of Behav and Exp Econ | 11769.1 | 86 | 348 | Applied micro |
| Journal of Asian Economics | 12425.8 | 87 | 357 | Applied micro |
| Review of Financial Economics | 11990.4 | 88 | 375 | Finance |
| Asian Economic Papers | 8884.4 | 89 | 398 | Applied micro |
| Computational Economics | 11908.6 | 90 | 411 | Micro |
| Journal of Economics of Ageing | 10555.7 | 91 | 463 | Applied micro |
| Southern Economic Journal | 12062.5 | 92 | 424 | General |
| Metroeconomica | 10766.9 | 93 | 471 | General |
| Journal of Cultural Econ | 10660.7 | 94 | 482 | Applied micro |
| Japanese Economic Review | 13731.0 | 95 | 442 | Applied micro |
| Pacific Economic Review | 9741.9 | 96 | 447 | Applied micro |
| Journal of Sports Economics | 8529.2 | 97 | 439 | Applied micro |
| Journal of Benefit-Cost Analysis | 8752.9 | 98 | 466 | Applied micro |
| Journal of Wine Economics | 8372.5 | 99 | 521 | Applied micro |
| Review of Quant Finance and Accounting | 14718.4 | 100 | 531 | Finance |
| Asian Econ and Fin Rev | 8114.1 | | | Others |
| Business and Econ Journal | 5274.7 | | | Others |
| Commodities | 9114.3 | | | Others |
| Econometrics | 9334.4 | | | Others |
| Economies | 11326.3 | | | Others |
| International J of Financial Studies | 12293.8 | | | Others |
| Journal of Applied Econ and Business | 5298.3 | | | Others |
| Journal of Risk and Financial Management | 11004.3 | | | Others |
| Regional Science and Env Econ | 10026.8 | | | Others |
| Social Sciences | 11145.4 | | | Others |

## Appendix B: Prompts used in each LLM

### 1. "Top 5 Journal" rating

'top5_journal_rating': "In your capacity as a reviewer for one of the most prestigious and highly selective top-5 economics journals (such as Econometrica, Journal of Political Economy, or The Quarterly Journal of Economics), please determine whether you would recommend this submission for publication using the following 6-point scale: 1 = Definite Reject: Fatal flaws in theory/methodology, insufficient contribution, or serious validity concerns that make the paper unsuitable for the journal, 2 = Reject with Option to Resubmit: Significant issues with theory, methodology, or contribution, but potentially salvageable with major revisions and fresh review, 3 = Major Revision: Substantial changes needed to theory, empirics, or exposition, but the core contribution is promising enough to warrant another round, 4 = Minor Revision: Generally strong paper with few small changes needed in exposition, robustness checks, or literature discussion, 5 = Very Minor Revision: Excellent contribution needing only technical corrections or minor clarifications, 6 = Accept As Is: Exceptional contribution ready for immediate publication",

### 0. Originality, Rigor, Scope, Impact, Written by AI ratings

"""Please evaluate the attached research according to the following criteria.

ORIGINALITY
"In your capacity as an editorial board/reviewer for this paper, please rate this paper's originality (0 = Completely unoriginal, …, 10 = Completely original)"

RIGOR
"In your capacity as an editorial board/reviewer for this paper, please rate this paper's rigor (0 = Not at all rigorous, …, 10 = Extremely rigorous)"

SCOPE
"In your capacity as an editorial board/reviewer for this paper, please rate this paper's scope (0 = Narrowest, …, 10 = Widest)"

IMPACT
"In your capacity as an editorial board/reviewer for this paper, please rate this paper's originality (0 = Minimal impact, …, 10 = Maximum impact)"

WRITTEN_BY_AI
"Please determine whether this paper was written by AI (0 = Definitely human-written, …, 10 = Definitely AI-generated)"
"""

# Appendix C: Binscatter Regressions: IMSE-optimal Binning

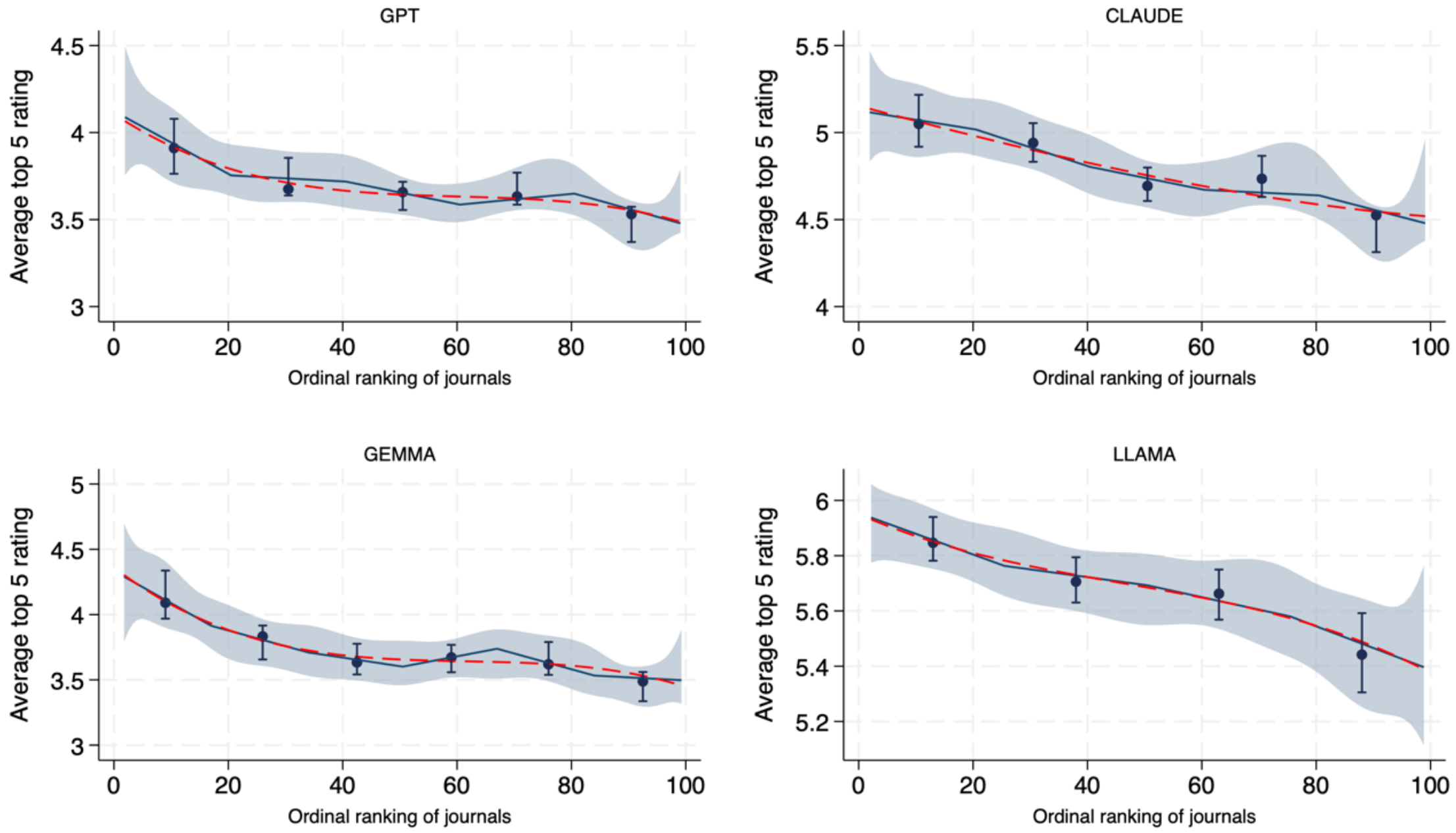

**C1:** Top 5 ratings and ordinal ranking of journals on RePEc (February 2025)

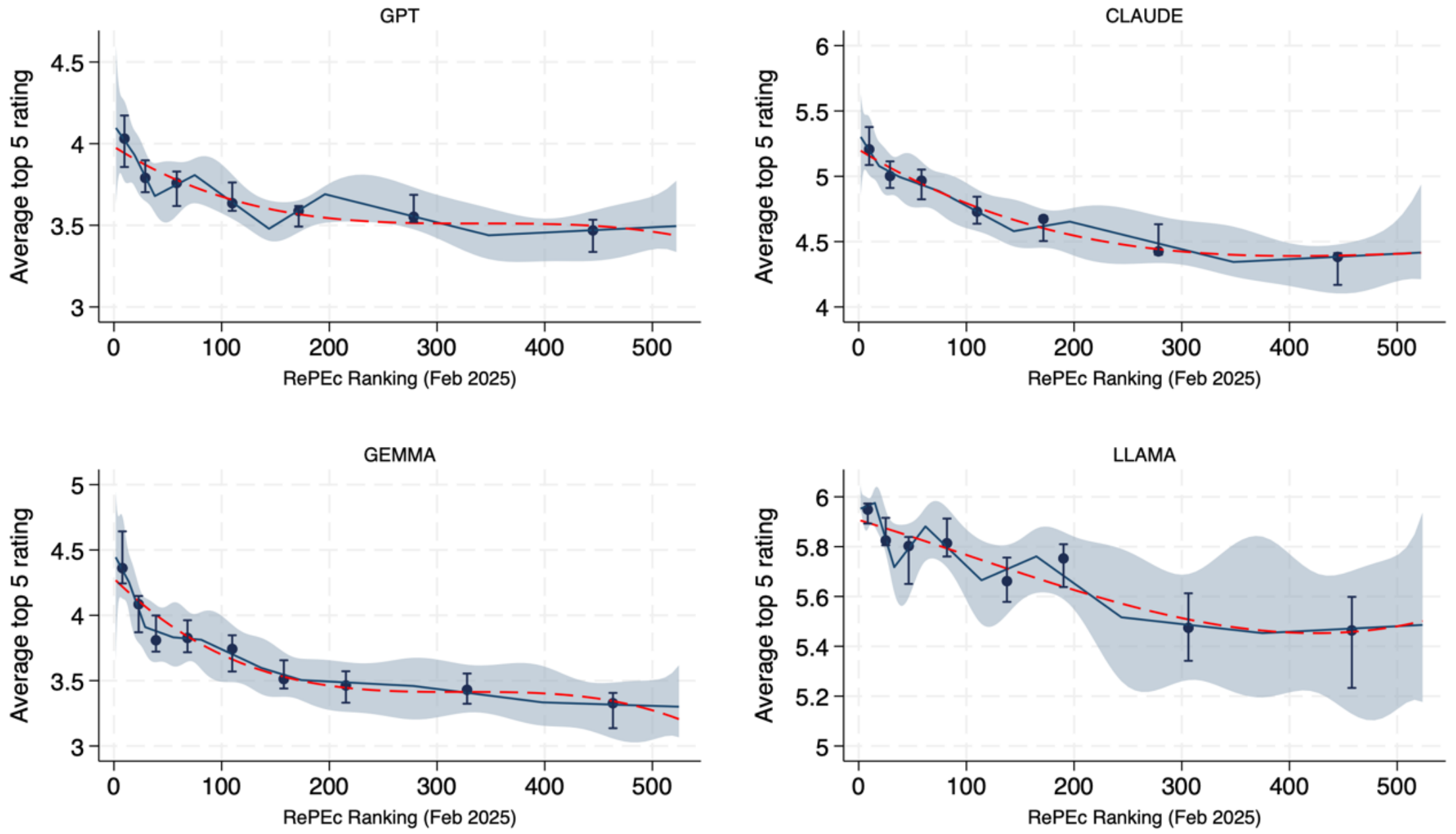

**C2*:*** Top 5 ratings and actual ranking of journals on RePEc (February 2025)

**Note:** Nonparametric estimates of LLM-assigned paper quality utilize IMSE-optimal binning (7 bins). This figure complements Figure 1 by estimating the same conditional expectation function, with the number of bins selected based on the integrated mean squared error (IMSE) rule. Confidence intervals (shaded) are constructed using a global cubic polynomial. The same control variables are as in Figure 1.

## Appendix D: OLS regressions on four domains of scholarly quality

| Variables | (1)<br>GPT<br>Top 5 rating | (2)<br>Claude<br>Top 5 rating | (3)<br>Gemma<br>Top 5 rating | (4)<br>LLaMA<br>Top 5 rating |
|---|---|---|---|---|
| GPT: originality | 0.371***<br>(0.045) | | | |
| GPT: rigor | 0.276***<br>(0.046) | | | |
| GPT: slope | 0.140***<br>(0.035) | | | |
| GPT: impact | 0.142***<br>(0.031) | | | |
| Claude: originality | | 0.451***<br>(0.042) | | |
| Claude: rigor | | 0.514***<br>(0.044) | | |
| Claude: scope | | 0.073**<br>(0.031) | | |
| Claude: impact | | 0.048*<br>(0.025) | | |
| Gemma: originality | | | 0.229***<br>(0.037) | |
| Gemma: rigor | | | 0.274***<br>(0.036) | |
| Gemma: scope | | | 0.107**<br>(0.049) | |
| Gemma: impact | | | -0.105**<br>(0.046) | |
| LLaMA: originality | | | | 0.607***<br>(0.233) |
| LLaMA: rigor | | | | 0.921*<br>(0.504) |
| LLaMA: scope | | | | -.007<br>(0.041) |
| LLaMA: impact | | | | 0.132***<br>(0.02) |
| Intercept | -3.795***<br>(0.470) | -4.533***<br>(0.407) | -0.071<br>(0.267) | -8.389***<br>(2.798) |
| Observations | 1220 | 1220 | 1220 | 1220 |
| $R^2$ | 0.225 | 0.281 | 0.251 | 0.205 |

*Notes: Standard errors are in parentheses. *** p<.01, ** p<.05, * p<.1*

## Appendix E: Coefficient plots by the LLM used to generate the fake AI papers

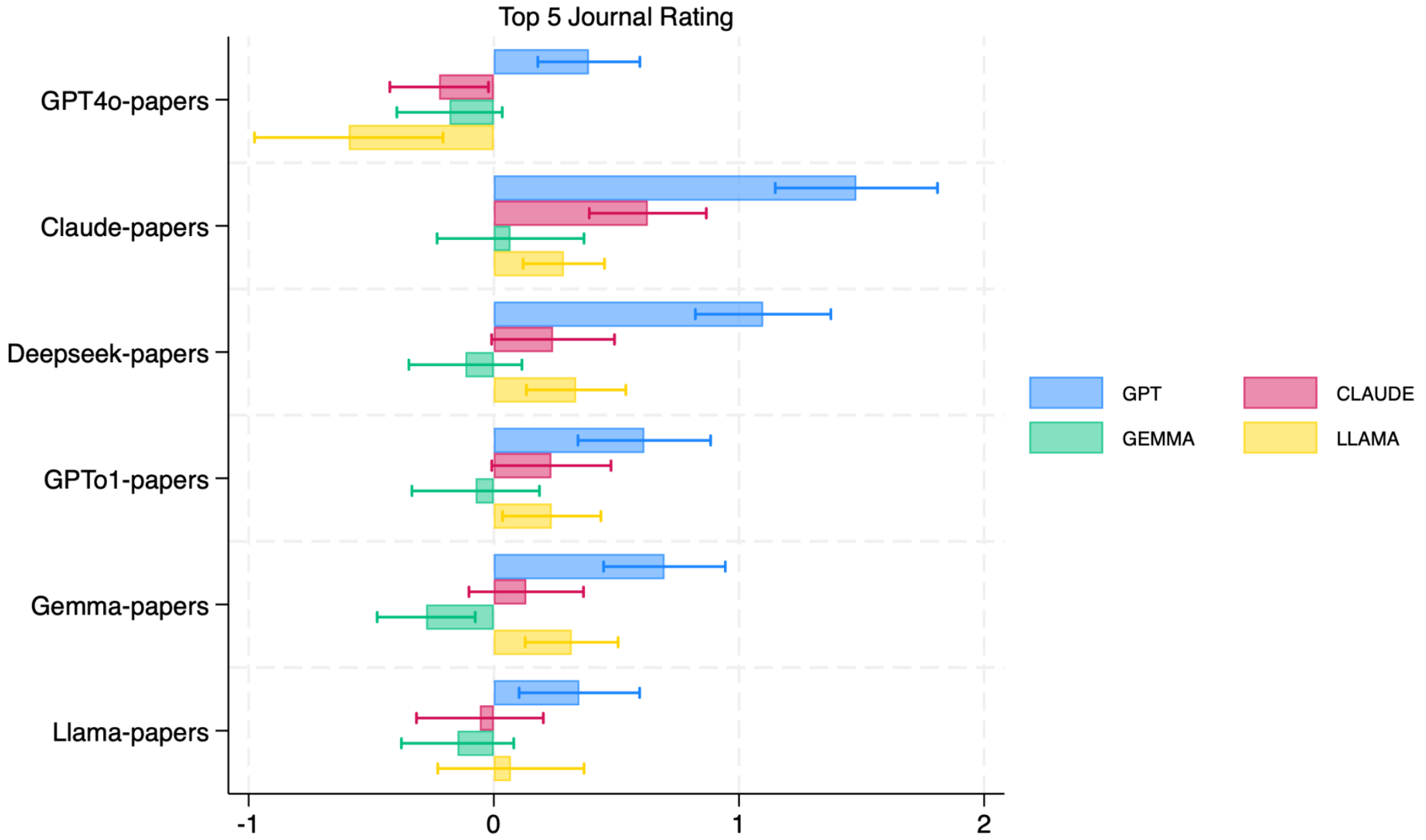

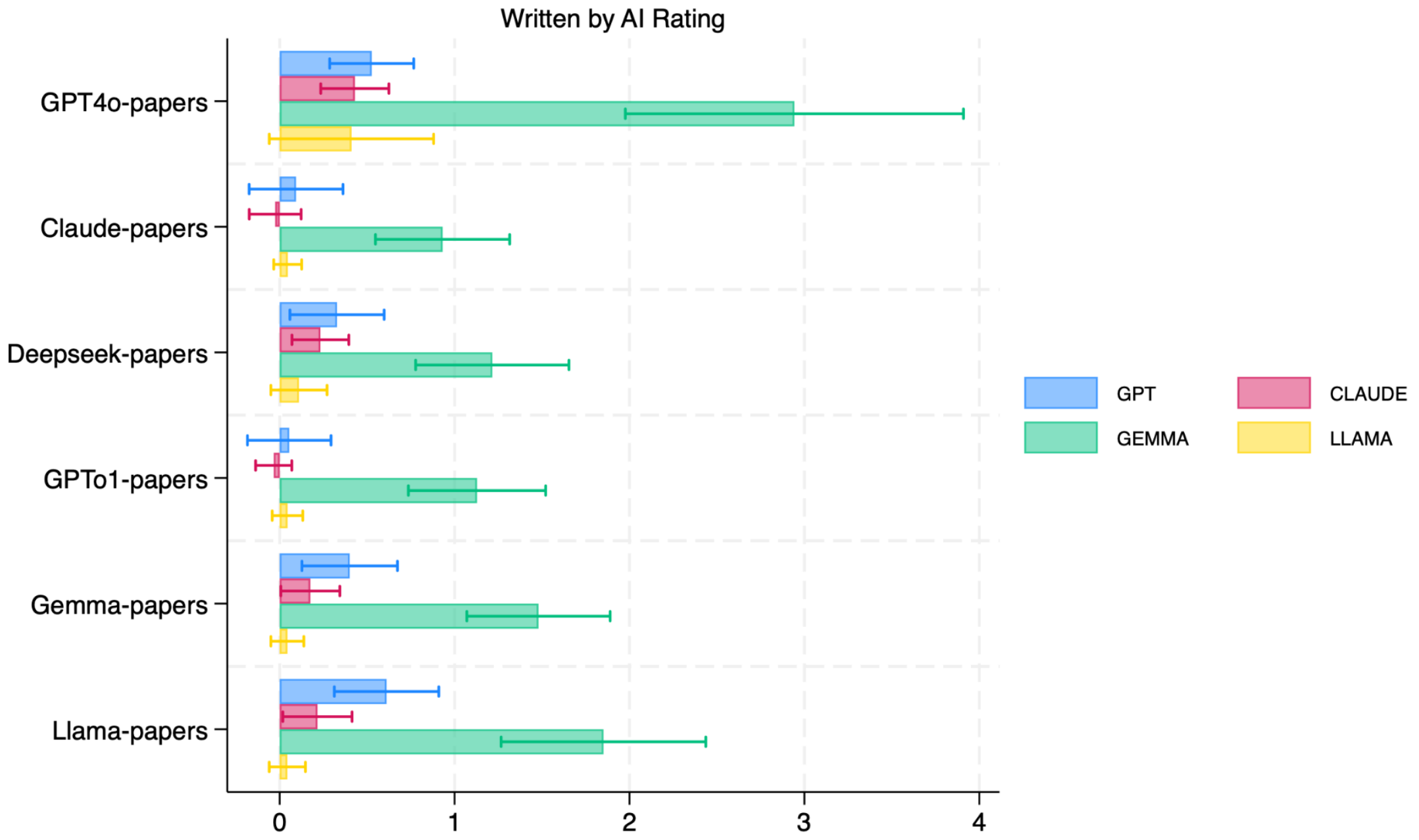

**Note:** Coefficient plots for each LLM model from regressions on the “Top 5 journal” rating (1–6 scale) and the “Written by AI” rating (1–10 scale), displaying only the coefficients of different LLMs used to generate the faked papers. The reference category is journals ranked 1st–25th. All regressions include controls for paper length, paper length squared, and field fixed effects. 95% confidence intervals are shown.